\documentclass{article}

\usepackage{arxiv}

\usepackage[utf8]{inputenc} % allow utf-8 input
\usepackage[T1]{fontenc}    % use 8-bit T1 fonts
\usepackage{hyperref}       % hyperlinks
\usepackage{url}            % simple URL typesetting
\usepackage{booktabs}       % professional-quality tables
\usepackage{amsfonts}       % blackboard math symbols
\usepackage{nicefrac}       % compact symbols for 1/2, etc.
\usepackage{microtype}      % microtypography
\usepackage{graphicx}
\usepackage{listings}
\usepackage{tabularx}
\usepackage{inconsolata}
\usepackage[tt=false]{libertine}
\usepackage{makecell}
\usepackage{float}
\usepackage{amsthm}
\usepackage{subcaption}
\usepackage{xcolor}
\usepackage[group-separator={,}]{siunitx}
\usepackage{glossaries}
\usepackage{doi}
\usepackage{caption}
\glsdisablehyper
\newacronym{IoT}{IoT}{Internet of Things}
\newacronym{OLLVM}{OLLVM}{Obfuscator-LLVM}
\newacronym{RF}{RF}{random forest}
\newacronym{MBA}{MBA}{mixed boolean-arithmetic}
\newacronym{CIL}{CIL}{C intermediate language}
\newacronym{CFG}{CFG}{control flow graph}
\newacronym{CNN}{CNN}{convolutional neural network}
\newacronym{SVM}{SVM}{support vector machine}
\newacronym{ASIC}{ASIC}{application-specific integrated circuit}
\newacronym{DFG}{DFG}{data flow graph}
\newacronym{DTW}{DTW}{dynamic time warping}
\newacronym{AST}{AST}{abstract syntax tree}
\newacronym{VM}{VM}{virtual machine}
\newacronym{WASI}{WASI}{WebAssembly System Interface}
\newacronym{WABT}{WABT}{WebAssembly Binary Toolkit}
\newacronym{ISA}{ISA}{instruction set architecture}

\definecolor{linkcolor}{HTML}{2A7CA5}

\hypersetup{
	colorlinks = true,
	linkcolor = linkcolor,
	citecolor = linkcolor,
	urlcolor = linkcolor,
}

\newtheoremstyle{defstyle}
{1em}    % Space above
{1em}    % Space below
{}       % Body font
{}       % Indent amount
{\bfseries}    % Theorem head font
{\normalfont.}    % Punctuation after theorem head     
{.5em}   % Space after theorem head
{}       % Theorem head spec

\theoremstyle{defstyle}
\newtheorem{definition}{Definition}

\newcommand{\inlinesection}[1]{\textbf{#1}.}

\title{Cryptic Bytes: WebAssembly Obfuscation\\ for Evading Cryptojacking Detection}

\author{ 
	\href{https://orcid.org/0009-0006-9239-1345}{\includegraphics[scale=0.06]{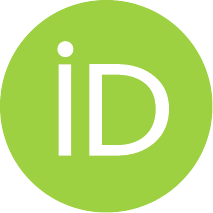}\hspace{1mm}Håkon Harnes} \\
	Department of Computer Science\\	
	Norwegian University of Science and Technology\\	
 Trondheim, Norway \\
	\texttt{haakaha@alumni.ntnu.no} \\
 \And%
\href{https://orcid.org/0009-0001-6072-4081}{\includegraphics[scale=0.06]{orcid.pdf}\hspace{1mm}Donn Morrison}\\
	Department of Computer Science\\
	Norwegian University of Science and Technology\\
	Trondheim, Norway\\
	\texttt{donn.morrison@ntnu.no} \\
}

\renewcommand{\shorttitle}{Cryptic Bytes: WebAssembly Obfuscation for Evading Cryptojacking Detection}

\hypersetup{
pdftitle={Cryptic Bytes: WebAssembly Obfuscation for Evading Cryptojacking Detection},
pdfsubject={q-bio.NC, q-bio.QM},
pdfauthor={Håkon Harnes, Donn Morrison},
pdfkeywords={WebAssmelby, obfuscation, cybersecurity, cryptojacking},
}

\begin{document}
\maketitle

\begin{abstract}
WebAssembly has gained significant traction as a high-performance, secure, and portable compilation target for the Web and beyond. However, its growing adoption has also introduced new security challenges. One such threat is cryptojacking, where websites mine cryptocurrencies on visitors' devices without their knowledge or consent, often through the use of WebAssembly. While detection methods have been proposed, research on circumventing them remains limited. In this paper, we present the most comprehensive evaluation of code obfuscation techniques for WebAssembly to date, assessing their effectiveness, detectability, and overhead across multiple abstraction levels. We obfuscate a diverse set of applications, including utilities, games, and crypto miners, using state-of-the-art obfuscation tools like Tigress and wasm-mutate, as well as our novel tool, emcc-obf. Our findings suggest that obfuscation can effectively produce dissimilar WebAssembly binaries, with Tigress proving most effective, followed by emcc-obf and wasm-mutate. The impact on the resulting native code is also significant, although the V8 engine's TurboFan optimizer can reduce native code size by 30\% on average. Notably, we find that obfuscation can successfully evade state-of-the-art cryptojacking detectors. Although obfuscation can introduce substantial performance overheads, we demonstrate how obfuscation can be used for evading detection with minimal overhead in real-world scenarios by strategically applying transformations. These insights are valuable for researchers, providing a foundation for developing more robust detection methods. Additionally, we make our dataset of over 20,000 obfuscated WebAssembly binaries and the emcc-obf tool publicly available to stimulate further research.
\end{abstract}


% ------------
% INTRODUCTION 
% ------------
\section{Introduction}

In 2015, Mozilla, Microsoft, Google, and Apple announced they were working on WebAssembly, a low-level bytecode language for the Web.
It allows high-level languages like C, C++, and Rust to be executed in the browser at near-native performance. 
In 2019, WebAssembly received formal recognition as a Web standard~\cite{w3c-standard}, marking a significant milestone as the first new language to gain native support on the Web alongside JavaScript in over two \textit{decades}.
Since then, it has gained widespread adoption by large corporations, like Google, Figma, and eBay, who have leveraged it to improve performance and port once desktop-only applications to the Web~\cite{earth-2021-google,online-2023-figma,ebay-2019-webassembly}.
Beyond the Web, WebAssembly has been extended to desktop applications~\cite{moller-2018-technical}, mobile devices~\cite{pop-2021-secure}, cloud computing~\cite{2022-fastlydocs}, blockchain virtual machines (VMs)~\cite{ewasm-ethereum-2022}, \gls{IoT}~\cite{liu-2021-aerogel}, embedded devices~\cite{scheidl-2020-valent}, and stand-alone runtimes~\cite{wasmer-online}.

However, the growing adoption of WebAssembly has also introduced new security challenges.
The most prominent example of this is cryptojacking, an attack strategy that exploits a website visitor's hardware resources to mine cryptocurrencies without their knowledge or consent.
The introduction of cryptocurrencies like Monero and VerusCoin, which can be mined using consumer-grade CPUs, has made cryptojacking a feasible attack vector~\cite{cpu-2023-mining}.
Although cryptojacking was initially implemented using JavaScript, WebAssembly has been the preferred method in recent years due to its superior performance.
The release of WebAssembly in 2017 was followed by a 459\% increase in cryptojacking incidents in 2018~\cite{cyberthreat-2018}.
By 2019, over half of all websites containing WebAssembly used it for cryptojacking~\cite{musch-2019-newkidweb}.
Reports from 2022 confirm the steady growth of cryptojacking, indicating that it remains a pervasive problem~\cite{kondratyev-2023-the}.

In response to the escalating threat of cryptojacking, numerous detection methods have been proposed.
Some methods rely entirely on static analysis~\cite{naseem-2021-minoslightweightreal, romano-2020-wasim, romano-2020-minerray}, while others use dynamic analysis to detect cryptojacking~\cite{wang-2018-seismicsecurelined, rodriguez-2018-rapid, kharraz-2019-outguarddetectingbrowser}.
Static methods leverage signature matching, control flow graph analysis, and neural networks, while dynamic methods focus on behavioral characteristics like CPU and memory usage, network traffic, and JavaScript events.

Surprisingly, there is limited research on how these detection methods can be side\-stepped.
Obfuscation, the process of making a program harder to analyze through applying various code transformations, has proven to be an effective evasion strategy for malware detection~\cite{park-2019-generation-evaluation, monirulsharif-2008-impedingmalwareanalysis}.
Although obfuscated WebAssembly binaries are a common occurrence on the Web~\cite{hilbig-2021-empiricalstudyreal}, the subject of WebAssembly obfuscation has been scarcely explored.
A short paper from 2021 touched upon this topic~\cite{bhansali-2022-firstlookcode}, and during our research, two papers emerged with a similarly narrow focus~\cite{cabrera-arteaga-2022-webassemblydiversificationmalware, Loose2023}, considering only a limited set of obfuscation and detection methods. 
This recent surge of interest underscores the timeliness and relevance of this issue and the evident gap in the literature.

The primary objective of this paper is to investigate and understand the application, effectiveness, and implications of code obfuscation for WebAssembly. 
We evaluate how well obfuscation can disguise the underlying nature of the code and evade cryptojacking detection. 
Moreover, we quantify the overhead introduced by obfuscation, both in terms of code size and hash rate, and assess whether the overhead is justifiable given the obfuscation advantages. 
Guided by these objectives, we investigate the following research questions:

\begin{itemize}
	\item \inlinesection{RQ1 -- Effectiveness} How effective are the transformations at obfuscating WebAssembly, and how is the resulting native code affected?
	\item \inlinesection{RQ2 -- Detectability} How effective is obfuscation at evading state-of-the-art cryptojacking detectors, and which transformations are the most effective?
	\item \inlinesection{RQ3 -- Overhead} To what extent do the transformations contribute to overhead in terms of code size and hash rate?
\end{itemize}

We extend the current body of literature with the following contributions:

\begin{itemize}
	\item  We perform a comprehensive evaluation of code obfuscation for WebAssembly across a diverse dataset, applying obfuscation at multiple abstraction levels -- a first in the literature. 
	\item We investigate how obfuscation can disguise the underlying code and evade state-of-the-art cryptojacking detectors.
	\item We develop and release emcc-obf,\footnote{\url{https://github.com/HakonHarnes/emcc-obf}} a novel obfuscation method for WebAssembly, and compare it with existing methods.
	\item  We quantify the overhead introduced by obfuscation, considering factors like code size and hash rate. We provide a granular analysis of the impact on different CryptoNight variants, a novelty in the existing body of literature.
	\item We compile a dataset of over 20,000 obfuscated WebAssembly binaries spanning diverse use cases, including all prominent CryptoNight variants, serving as a significant resource for future studies.\footnote{\url{https://github.com/HakonHarnes/wasm-obf/releases/tag/v1.0}}
\end{itemize}

% ----------
% BACKGROUND 
% ----------

\section{Background} \label{ch:background}

In this section, we provide an overview of the key concepts and technologies relevant to our research. We begin by discussing WebAssembly, its characteristics, and its growing adoption across various platforms. We then explore the threat of cryptojacking and the analysis techniques used to detect it, focusing on static methods. Finally, we delve into the concept of obfuscation and its potential for evading detection.

\subsection{WebAssembly}
\label{sec:webassembly}

\begin{figure}[ht]
	\begin{center}
		\includegraphics[width=150mm, height=100mm, keepaspectratio]{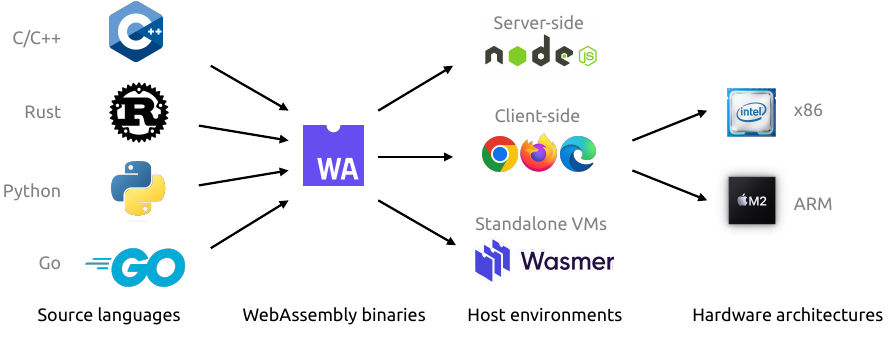}
		\caption{WebAssembly serves as an intermediate bytecode, bridging the gap between multiple source languages and host environments. The host environments compile the WebAssembly binaries into native code for underlying hardware architectures.}
		\label{fig:wasm-overview}
	\end{center}
\end{figure}

\inlinesection{Overview}
WebAssembly is a low-level bytecode language designed for the Web.
It acts as an intermediate step, enabling interoperability between source languages and hardware architectures, as depicted in Figure \ref{fig:wasm-overview}. 
As a compilation target, WebAssembly provides a universal language into which different types of source code can be compiled and subsequently executed in various host environments. 
For example, programs written in Rust (a source language) can be compiled to WebAssembly and run in a browser using the V8 engine (a host environment), which then compiles the WebAssembly code into native code specific to the underlying hardware architecture.

\inlinesection{Modules}
WebAssembly modules are the fundamental units for deployment, loading, and compilation.
These modules contain definitions for various components, including types, functions, tables, memories, and globals.
Each function defined within the module is associated with a corresponding function body in the code section.
The module can export its components, allowing the host environment to import and utilize them as needed.

\inlinesection{Text and Binary Format}
WebAssembly modules can be represented in two formats; the binary format (\texttt{wasm}) and the human-readable text format (\texttt{wat}).
The binary format is designed to be compact, enabling efficient network transmission and parsing, and is typically generated by compilers and instantiated in runtimes.
The human-readable text format serves a different purpose, being designed for debugging, testing, and occasional manual editing, akin to native assembly languages.
It is possible to convert between these formats using tools like \texttt{wasm2wat} and \texttt{wat2wasm}, which are part of the \gls{WABT}.

\inlinesection{Stack and Variables}
WebAssembly modules are executed on a stack-based \gls{VM}.
In this system, instructions operate by popping their inputs from and pushing their results to an implicit evaluation stack.
There are no registers in this system; instead, values are stored in local or global variables.
Global variables are visible to the entire module, while local variables are only accessible within the current function.
The evaluation stack, local variables, and global variables are managed by the \gls{VM}.

\inlinesection{Host Environment}
WebAssembly modules are executed within a host environment, which provides the necessary functionality for the module to perform actions such as file or network access. 
In a browser, the host environment is typically provided by the JavaScript engine, such as V8 or SpiderMonkey.
The WebAssembly-JavaScript API allows WebAssembly exports to be wrapped in JavaScript functions, enabling them to be called from JavaScript code. 
Conversely, WebAssembly code can import and call JavaScript functions, enabling bidirectional communication between the two languages.
Beyond the browser, WebAssembly modules can be executed in other host environments, including server-side environments like Node.js and stand-alone \gls{VM}s like Wasmer. 
These host environments provide their own APIs for WebAssembly modules to interact with, giving them access platform-specific features and resources. 
For example, modules running on stand-alone \gls{VM}s can leverage the \gls{WASI} to interact with the file system.

\inlinesection{Source Languages}
WebAssembly's low-level nature makes it an ideal compilation target for systems languages like C, C++, and Rust.
Both the Clang and the Rust compilers have native support for WebAssembly, enabling direct generation of WebAssembly modules~\cite{clang-2023-wasm, rust-2023-webassembly}.
Moreover, Emscripten, a toolchain built on Clang and LLVM, is capable of porting C and C++ code to the Web using WebAssembly~\cite{emscripten-page}.
In addition to compiling C and C++ code to WebAssembly, Emscripten also produces the corresponding JavaScript ``glue'' code.
The JavaScript code is responsible for module instantiation and providing the necessary functionality for it to interact with the host environment.
For example, it pipes the output from the \texttt{printf} function to the browser's console.
Similarly, wasm-bindgen enables high-level interactions between Rust and JavaScript for use in the browser~\cite{wasm-bindgen}.
Even garbage-collected languages like C\#, Python, and more recently Kotlin and Dart, can be compiled to WebAssembly.

\inlinesection{V8 Compilation}
The V8 engine, which is used in Google Chrome, employs a two-tiered WebAssembly compilation pipeline~\cite{v8-compiler}.
Initially, the baseline Liftoff compiler lazily compiles functions when they are first called.
That is to say, if a function is never called, it is never compiled to native code.
Liftoff iterates over the WebAssembly code just once and immediately emits native code, which allows for fast code generation, albeit with a limited set of optimizations.
Once Liftoff compilation is done, the native code is registered with the WebAssembly module for immediate future use.
For functions that are frequently invoked, termed ``hot'' functions, the V8 engine uses its optimizing TurboFan compiler.
Unlike Liftoff, TurboFan is a multi-pass compiler that constructs multiple internal representations of the code during compilation, enabling advanced optimizations.
When a function is deemed hot, TurboFan is triggered to recompile and optimize it in the background.
The resulting optimized native code then replaces the existing Liftoff-generated code, ensuring increased performance for all subsequent calls to that function.

\subsection{Cryptojacking}
\label{sec:drive-by-mining}

\begin{figure}[ht]
	\begin{center}
		\includegraphics[width=0.8\textwidth]{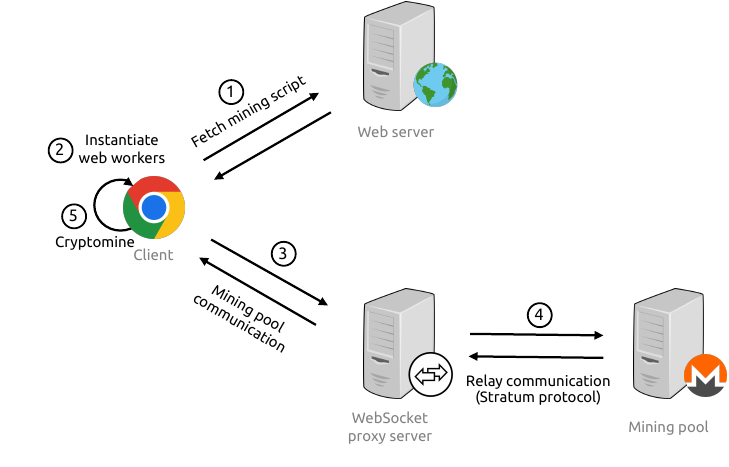}
	\end{center}
	\caption{Cryptojacking process: The mining script is fetched from the web server, which instantiates the web workers and connects to the WebSocket proxy server. The proxy server relays the communication back to the mining pool.}
	\label{fig:drive-by-mining}
\end{figure}

Cryptojacking\footnote{Also referred to as drive-by mining.} is a type of attack that involves using the hardware resources of a website visitor to mine cryptocurrencies without their knowledge or consent.
Such an approach has become feasible with the advent of cryptocurrencies like Monero~\cite{monero-2023-the} and VerusCoin~\cite{verus-2023-verus}, which can be mined using consumer-grade CPUs~\cite{cpu-2023-mining}.
Cryptojacking can be executed through self-hosted mining or by compromising web servers through software vulnerabilities or misconfigurations~\cite{thomson-2018-pulitzer,odonnell-2018-cryptojacking}, and then subsequently installing the mining scripts on the compromised web servers.
Alternatively, mining scripts can be distributed through advertising platforms~\cite{aol-2018-cryptocurrency}, compromised third-party libraries integrated within various websites~\cite{williams-2018-uk}, or through adversarial Docker images~\cite{chickowski-2022-container}.
Interestingly, browser-based crypto mining has been used as an alternative income stream by organizations like UNICEF~\cite{unicef-mining-2018}, although with user approval, differentiating it from adversarial cryptojacking.

Initially implemented with JavaScript only and popularized with the launch of CoinHive in 2017, WebAssembly has become the preferred method for cryptojacking in recent years due to its superior performance.
The release of WebAssembly in 2017 was followed by a 459\% increase in cryptojacking incidents in 2018~\cite{cyberthreat-2018}.
By 2019, more than half of all websites containing WebAssembly used it for cryptojacking~\cite{musch-2019-newkidweb}.
Although CoinHive shut down in 2019~\cite{porter-2019-popular}, reports from the first three quarters of 2022 confirm the steady growth of cryptojacking~\cite{kondratyev-2023-the}, indicating that it remains a pervasive problem.

Today, cryptojacking is implemented with a combination of JavaScript and WebAssembly.
The JavaScript code is responsible for coordinating the mining process by communicating with the mining pool, while WebAssembly is used to calculate the hashes.
The process is depicted in Figure \ref{fig:drive-by-mining}.
First, (1) when a user visits a website, the JavaScript and WebAssembly code is fetched from the web server.
Then, (2) the JavaScript code checks how many CPU cores are available on the host machine and spawns web workers, one for each available core.
Each web worker instantiates the WebAssembly module and (3) connects to the WebSocket proxy server.
The WebSocket proxy server (4) connects to the mining pool and retrieves the mining job.
Lastly, the communication is relayed back to the web workers, which (5) calculate the hashes and send the results back to the mining pool through the proxy server.
The communication between the web workers and the mining pool is usually implemented using the Stratum protocol~\cite{online-2023-stratumv2}.

\subsection{Analysis Techniques}
\label{sec:analysis-techniques}

Analysis techniques can be used in a variety of ways, not only for detecting malware but also for vulnerability identification~\cite{lopes-2021-discoveringvulnerabilitieswebassembly}, performance optimization~\cite{thiel-2023-an}, and for understanding and debugging code~\cite{nethercote2007valgrind}.
Although different in their objective, these techniques often follow similar methodologies.
As such, the analysis techniques, including those evaluated in this paper, can be classified along the following dimensions:

\inlinesection{Static and Dynamic}
Static analysis examines the program without executing it, enabling fast, real-time detection. 
However, its precision often falls short of dynamic analysis due to its reliance on approximating the actual runtime behavior~\cite{damodaran-2015-comparisonstaticdynamic}.
Moreover, obfuscation has been found to be effective against static analysis~\cite{moser-2007-limitsstaticanalysis}.
 In contrast, dynamic analysis observes the behavior of the program during execution, typically providing higher accuracy than static analysis~\cite{damodaran-2015-comparisonstaticdynamic}.
Although advantageous, it can be resource-intensive and time-consuming, and it may struggle to explore all possible program behaviors due to a large or potentially infinite search space~\cite{ye-2017-surveymalwaredetection}.
Although generally more resilient to obfuscation than static methods, the presence of obfuscation can increase time consumption and still lead to a considerable false positive rate~\cite{bazrafshan-2013-surveyheuristicmalware}.

\inlinesection{Rule-Based and Machine Learning}
Rule-based methods evaluate programs according to a set of predefined rules or patterns, offering transparent and comprehensible decision-making.
These methods can provide high accuracy, provided the rules are precisely formulated.
However, their effectiveness can be undermined by novel threats or obfuscation techniques that circumvent established rules~\cite{monirulsharif-2008-impedingmalwareanalysis}.
On the other end of the spectrum, machine learning models are trained on large datasets, where each entry is annotated with the expected output.
After training, the models apply the learned patterns to predict outcomes on previously unseen instances.
Although machine learning methods generally handle obfuscation better than rule-based methods, they are still vulnerable to adversarial attacks.
These attacks involve making subtle modifications to input data with the intent to deceive the model, thereby reducing its accuracy~\cite{szegedy-2013-intriguingpropertiesneural, goodfellow-2014-explainingharnessingadversarial, Loose2023}.

\subsubsection{Detecting Cryptojacking}
\label{sec:background-detecting-drive-by-mining}

To address the increasing threat of cryptojacking, a number of analysis techniques have been proposed.
The literature presents methods based on both static~\cite{naseem-2021-minoslightweightreal, romano-2020-wasim, romano-2020-minerray} and dynamic analysis~\cite{bian-2020-minethrottledefendingwasm, rodriguez-2018-rapid, kharraz-2019-outguarddetectingbrowser}, using both rule-based and machine learning-based approaches.
Static methods use a wide range of techniques, including signature matching, \gls{CFG} analysis, and neural networks.
Dynamic methods rely on behavioral characteristics such as CPU and memory usage, network traffic, and JavaScript events.

In this paper, we focus on static analysis techniques for several reasons.
Primarily, dynamic analysis proves impractical due to its substantial overhead,  ranging from 40\% to 100\%~\cite{rodriguez-2018-rapid, wang-2018-seismicsecurelined}.
This would likely degrade the user experience and, therefore, it is not a viable option for real-world applications.
Although dynamic analysis can be used for offline analysis, it is not a feasible solution either, as the blacklists need to be updated frequently and can be circumvented through diversification~\cite{cabrera-arteaga-2022-webassemblydiversificationmalware}.
Moreover, several dynamic methods depend on platform or browser-centric features, such as the MessageLoop event~\cite{kharraz-2019-outguarddetectingbrowser}, or Chrome debugging features~\cite{rodriguez-2018-rapid}, rendering it difficult to implement in real-world applications.
Lastly, empirical evidence demonstrates that static methods are just as effective as dynamic methods in combating cryptojacking, with F$_1$ scores ranging from 0.95~\cite{naseem-2021-minoslightweightreal} to 1.00~\cite{romano-2020-minerray}, compared to dynamic methods scoring between 0.96~\cite{rodriguez-2018-rapid} and 0.98~\cite{wang-2018-seismicsecurelined}.

\begin{figure}[H]
	\begin{center}
		\includegraphics[width=1.0\textwidth, keepaspectratio]{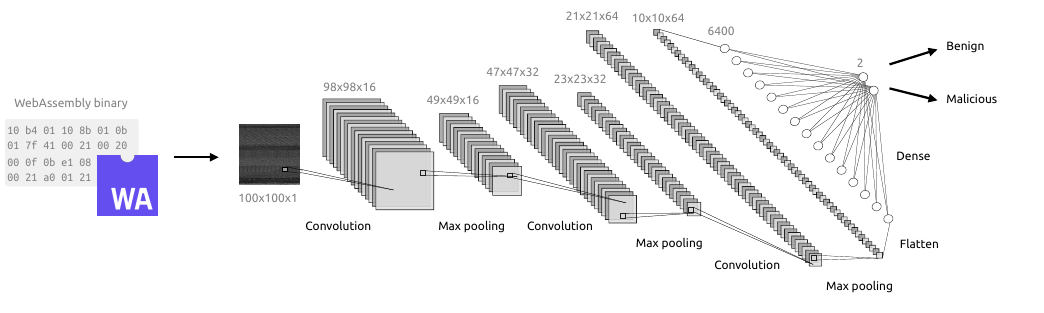}
		\caption{Overview of MINOS: The WebAssembly binary is converted to a grayscale image and fed to the MINOS network. The network predicts whether the binary is benign or malicious.}
		\label{fig:minos-network}
	\end{center}
\end{figure}

\inlinesection{MINOS}
MINOS~\cite{naseem-2021-minoslightweightreal} is a machine learning-based method that uses an image-based classification deep learning approach to identify cryptojacking. As illustrated in Figure \ref{fig:minos-network}, the WebAssembly binary is first converted into a 100x100 grayscale image. This image is then used as input to a \gls{CNN}, which has been trained on a dataset of malicious and benign WebAssembly binaries. The \gls{CNN} attempts to determine whether the WebAssembly binary performs cryptojacking based on the patterns it observes in the grayscale image. The model was able to achieve an F$_1$ score of 0.95 with an average detection time of just 25.9 milliseconds.

\begin{figure}[h]
	\begin{center}
		\includegraphics[width=1.0\textwidth, keepaspectratio]{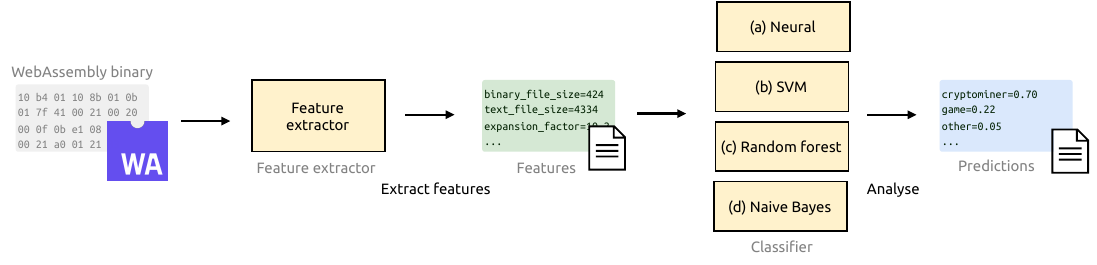}
		\caption{Overview of WASim: Features are extracted from the WebAssembly binaries and fed into a classifier. The classifier model is either: (a) Neural, (b) SVM, (c) Random forest, or (d) Naive Bayes. The classifier outputs a usage report containing the predictions.}
		\label{fig:wasim-overview}
	\end{center}
\end{figure}

\inlinesection{WASim}
WASim~\cite{romano-2020-wasim} is a machine learning-based method that extracts and analyzes a set of features for detecting cryptojacking.
The procedure is depicted in Figure \ref{fig:wasim-overview}.
First, the WebAssembly binary is converted from the binary format (\texttt{wasm}) to the text format (\texttt{wat}), which is then parsed to extract a set of features.
Then, these features are used by the classifier models to predict the use case of the WebAssembly module, for example, classifying it as a crypto miner, game, or other application.
The authors implemented several classifier models, including neural, \gls{SVM}, \gls{RF}, and naive Bayes, achieving accuracies of 91.6\%, 87\%, 82\%, and 64\%, respectively.

\begin{figure}[H]
	\begin{center}
		\includegraphics[width=1.0\textwidth, keepaspectratio]{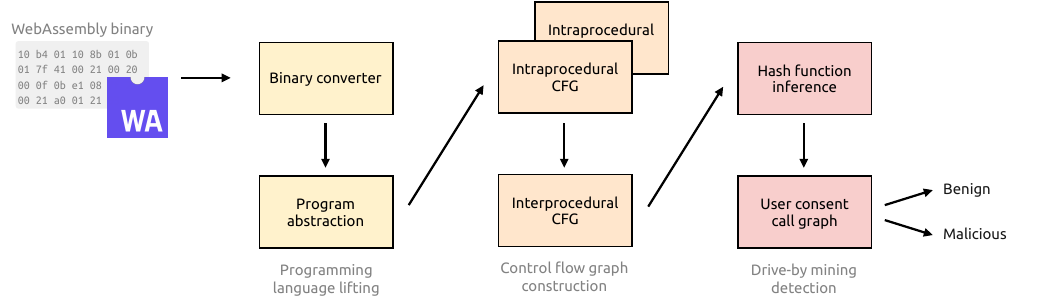}
		\caption{Overview of MinerRay: The WebAssembly binary is converted into a custom intermediate language, from which an interprocedural CFG is constructed. The control flow is then analyzed to detect cryptojacking, as well as for checking user consent.}
		\label{fig:minerray-overview}
	\end{center}
\end{figure}

\inlinesection{MinerRay}
MinerRay~\cite{romano-2020-minerray} is a rule-based method that analyzes the control flow of the WebAssembly module to detect cryptojacking.
As illustrated in Figure \ref{fig:minerray-overview}, the process is divided into three parts: Programming language lifting, \gls{CFG} construction, and cryptojacking detection.
For programming language lifting, it uses a set of abstraction rules to translate the WebAssembly opcodes to a custom intermediate representation.
Given the intermediate representation, an intraprocedural \gls{CFG} is constructed for each function.
These intraprocedural CFGs are then linked together to create an interprocedural CFG that represents the entire program.
MinerRay uses this interprocedural CFG to identify potential hashing algorithms by analyzing the control flow of the program and looking for patterns that match the semantics of hashing functions. To determine whether the user is informed about crypto mining, MinerRay employs a dynamic approach that explores \texttt{onclick} events of HTML objects. It checks if the \texttt{onlick} events can trigger WebAssembly APIs, such as \texttt{WebAssembly.instantiate}.
MinerRay achieved an F$_1$ score of 0.99 with an average detection time of 1.9 seconds.

\inlinesection{VirusTotal}
VirusTotal~\cite{virustotal-2022} uses an extensive set of antivirus scanners to detect malware.
Of the 70 antivirus scanners, 59 are able to scan WebAssembly binaries, including prominent ones such as AVG, Avast, and McAfee.
Each antivirus scanner integrated within the system incorporates distinct heuristic methods tailored for the detection of specific types of malware.
In the literature, it has been used to detect cryptojacking~\cite{hilbig-2021-empiricalstudyreal, wang-2018-seismicsecurelined, cabrera-arteaga-2022-webassemblydiversificationmalware}.

\subsection{Obfuscation}
\label{sec:obfuscation}

Obfuscation involves transforming a given program into one that is syntactically different but semantically equivalent~\cite{nagra2009surreptitious}.
It has been used for a variety of purposes, including prevention of reverse engineering~\cite{blazy-2015-datataintingobfuscation}, prevention of software modification~\cite{ghosh-2015-matryoshkastrengtheningsoftware}, hiding static data~\cite{kanzaki-2015-pinpointinghidingsurprising}, and for malware evasion~\cite{monirulsharif-2008-impedingmalwareanalysis, park-2019-generation-evaluation}.
The obfuscation process involves the application of a set of code transformations, which can be formally defined as follows:

\begin{definition}[Transformation]
	\label{def:transformation}
	Let $P \xrightarrow{T} P'$ be a transformation of a program $P$ to a program $P'$. The transformation is an obfuscating transformation if $P$ and $P'$ have the same observable behaviour but are different syntactically~\cite{collberg-1997-taxonomyobfuscatingtransformations}.
\end{definition}

The transformations applied in this paper, aligned with the taxonomy proposed by Collberg et al.~\cite{collberg-1997-taxonomyobfuscatingtransformations}, are categorized as follows:

\begin{itemize}
	\item \inlinesection{Control Obfuscation} Involves manipulating a program's control flow through techniques like adding false conditional statements and altering loop conditions.
	\item \inlinesection{Data Obfuscation} Alters the representation of data structures and values within the code, utilizing methods like splitting data structures and using complex encodings, while maintaining their semantic meaning.
	\item \inlinesection{Preventive Transformations} Designed to thwart specific types of code analysis or reverse engineering tools, incorporating techniques like anti-debugging, anti-tampering, and encoding mechanisms.
	\item \inlinesection{Layout Obfuscation} Modifies the arrangement of code elements to disrupt readability and traceability, including the removal of source code formatting and reordering of instruction sequences.
\end{itemize}

\subsubsection{Diversification}
Diversification is a technique that generates multiple distinct yet semantically equivalent versions of a program, with an emphasis on enhancing resilience against attacks rather than obscuring the analysis or understanding of the code.
Unlike obfusation, which primarily aims to obscure code analysis, diversification focuses on creating multiple program variants, ensuring that a single exploit cannot compromise all instances.
While diversification can inadvertently lead to code obfuscation, its primary goal is to improve security through variability.
In the context of this discussion, the terms \textit{mutations} and \textit{transformations} are used interchangeably to refer to either obfuscating transformations or diversifying mutations.

\subsubsection{Obfuscation Tools}

There are no known obfuscation tools that operate at the WebAssembly level.
However, the wasm-mutate~\cite{bytecodealliance-2023-wasm0mutate} diversifier can be used to diversify WebAssembly binaries, which may inadvertently lead to obfuscation.
Another option is to obfuscate code at a higher abstraction level, such as source code or LLVM bitcode, and then subsequently compile the obfuscated code to WebAssembly.
In this paper we apply obfuscation at multiple abstraction levels, using the following tools:

\begin{itemize}
	\item \inlinesection{Tigress} Tigress~\cite{tigress-2023-home} is a source-to-source obfuscator for C written in OCaml. It has been thoroughly evaluated in the literature~\cite{bhansali-2022-firstlookcode, banescu-2016-codeobfuscationsymbolic}, and has been found to be successful in evading cryptojacking detection~\cite{bhansali-2022-firstlookcode}.
	\item \inlinesection{OLLVM} \gls{OLLVM}~\cite{junod-2015-ollvm} is implemented as middle-end passes in the LLVM compilation suite. It has been used for preventing reverse engineering~\cite{banescu-2016-codeobfuscationsymbolic, lim-2017-antireverseengineering}, but not for evading cryptojacking detection. In this paper, we develop emcc-obf for applying \gls{OLLVM} obfuscation to WebAssembly.
	\item \inlinesection{wasm-mutate} wasm-mutate~\cite{bytecodealliance-2023-wasm0mutate} is a wasm-to-wasm diversifier written in Rust. Wasm-mutate has been used for fuzzing and has been found to be successful in evading cryptojacking detection~\cite{arteaga2022wasm, cabrera-arteaga-2022-webassemblydiversificationmalware}.
\end{itemize}

\subsection{Code Similarity}
\label{sec:code-similarity}

Traditional code similarity evaluation methods, like cyclomatic complexity~\cite{cyclomatic-1976}, Halstead complexity measures~\cite{halstead1977elements}, and lines of code, focus on structural aspects of code such as control flow, operator usage, and code size.
However, these metrics are limited as they fail to capture the nuanced semantics of code and are vulnerable to superficial changes like code refactoring.
These shortcomings are particularly evident in the context of WebAssembly, as it is a stack-based language where the order of instructions affects the semantics of the program, a factor these traditional metrics do not consider.

Sequence alignment methods take into account the order of instructions, making them well-suited for stack-based languages like WebAssembly.
Among these methods, \gls{DTW} has proven to be more accurate than other sequence alignment algorithms~\cite{aach2001aligning}, and it has been successfully used for aligning and comparing stack traces and WebAssembly binaries~\cite{arteaga-2019-scalablecomparisonjavascript,arteaga-2020-crowcodediversification}.
The fundamental concept of \gls{DTW} is to identify the optimal alignment between two sequences.
The process aims to minimize the sum of absolute differences, commonly referred to as the \textit{distance}, between corresponding elements within the sequences.
\gls{DTW} computes a cost matrix representing the pairwise distances between all possible pairs of points in the two sequences.
the goal is to find a path through this cost matrix, the so-called \textit{warping path}, which minimizes the total cumulative distance.
To measure code similarity, we represent WebAssembly binaries as sequences of instructions and compare them using \gls{DTW}.

% ------------
% RELATED WORK 
% ------------

\section{Related Work}
\label{ch:related-work}

In this section, we review the existing literature on WebAssembly obfuscation and its application in evading cryptojacking detection. We discuss the recent studies that have explored the use of obfuscation techniques such as Tigress, wasm-mutate, and binary manipulation to disguise WebAssembly binaries and bypass state-of-the-art detectors.

Bhansali et al.~\cite{bhansali-2022-firstlookcode} used Tigress to obfuscate C source code before compiling it to WebAssembly. They found they could successfully evade detection from MINOS.
However, they did not ensure the WebAssembly binaries were still functional after obfuscation. 
The overhead caused by obfuscation was not measured, either.
Cabrera Arteaga et al.~\cite{cabrera-arteaga-2022-webassemblydiversificationmalware} demonstrated how wasm-mutate can be utilized to diversify WebAssembly binaries, effectively evading both MINOS and VirusTotal with minimal performance overhead. 
Loose et al.~\cite{Loose2023} proposed a novel technique that employs binary manipulation through instrumentation to incorporate adversarial examples into code sections within WebAssembly modules, successfully evading MINOS. 
However, they measured the performance overhead using a generic SHA256-hashing library instead of crypto mining binaries, which may limit the validity of their results.

In a similar vein, studies have turned to WebAssembly as a means of obfuscating JavaScript.
Romano et al.\cite{romano-2022-wobfuscatorobfuscatingjavascript} proposed Wobfuscator, a technique based on a set of code transformations that opportunistically translates specific parts of JavaScript code into WebAssembly. 
Similarly, Wang et al.\cite{wang-2019-leveragingwebassemblynumerical} introduced JSPro, a tool that converts JavaScript into WebAssembly for obfuscation purposes.
It is important to clarify that their objective diverges from ours; they seek to obfuscate JavaScript code by using WebAssembly, while we concentrate on obfuscating the WebAssembly code itself.

% -----------
% METHODOLOGY 
% -----------

\section{Methodology} \label{ch:methodology}

In this section, we describe the methodology employed in our research. We start by detailing the experimental setup, including the system configuration and dataset used. We then discuss the implementation of the obfuscation methods, cryptojacking detectors, and the dynamic time warping algorithm used for comparing WebAssembly binaries. Finally, we define the evaluation metrics used to assess the effectiveness, detectability, and overhead of the obfuscation methods.

\newpage
\subsection{Experimental Setup}
\label{sec:experimental-setup}

The experiments were performed on a \gls{VM} running Debian 10 with kernel 4.19.0-22, equipped with an AMD EPYC 7742 CPU running at 2.24 GHz. 
Although the VM had 64 CPU cores available, the experiments relating to cryptojacking were performed using only 4 cores to ensure that the results would be comparable with consumer-grade hardware and mobile devices, which typically have fewer cores. 
To guarantee the reproducibility of the experiments across various system configurations, the code has been containerized using Docker.

\subsubsection{Dataset}

\begin{table}[t]
	\small
	\begin{center}
            \begin{tabular}{lll}
			\toprule
			Category      & Name             & Description                                                          \\
			\midrule
			Utilities     & lcs            & Calculates the longest common subsequence of two strings             \\
			              & tree             & Lists folder contents in a tree format                               \\
			              & wgsim            & Whole-genome simulation tool for generating sequencing reads         \\
			              & seqtk            & Toolkit for processing sequences in FASTA/Q formats                  \\
			              & smith-waterman   & Algorithm for local sequence alignment of protein and DNA            \\
			              & needleman-wunsch & Algorithm for global sequence alignment of protein and DNA           \\
			\midrule
			Games         & pong             & Arcade game simulating table tennis                                  \\
			              & snake            & Arcade game where the player controls a snake                        \\
			              & f1-race          & Racing game that simulates Formula 1 car races                       \\
			              & breakout         & Arcade game where you control a paddle to hit a bouncing ball        \\
			              & game-of-life     & Cellular automaton simulating the evolution of cells                 \\
			              & wasm-asteroids   & Arcade game where you control a spaceship to shoot asteroids         \\
			\midrule
			Crypto miners & cn-0             & Variant 0 -- Original CryptoNight algorithm                          \\
			              & cn-1             & Variant 1 -- Also known as Monero v7                                 \\
			              & cn-2             & Variant 2 -- ASIC-resistant version                                  \\
			              & cn-r             & Variant 4 -- Also known as CryptoNightR                              \\
			              & cn-lite-0        & Lite variant 0 -- cn-0 with half the memory and iterations           \\
			              & cn-lite-1        & Lite variant 1 -- cn-1 with half the memory and iterations           \\
			              & cn-lite-2        & Lite variant 2 -- cn-2 with half the memory and iterations           \\
			              & cn-half          & Half variant -- cn-2 with half the iterations                        \\
			              & cn-rwz           & Reduced work variant -- cn-2 with a quarter the iterations           \\
			              & cn-pico-trtl     & Pico turtle variant -- cn-2 with an eighth the memory and iterations \\
			\bottomrule
		\end{tabular}
		\caption{Dataset used in this paper, spanning a wide range of categories, including utilities, games, and crypto miners.}
		\label{tab:dataset}
	\end{center}
\end{table}

Detailed in Table \ref{tab:dataset}, the dataset used in this paper covers a broad spectrum of categories, including utilities, games, and crypto miners.
All the applications are open-source projects written in C.
The utilities were carefully selected to represent a wide range of functionality, featuring Linux utilities, sequence alignment algorithms, and simulations.
The games were chosen to represent varying levels of complexity, from classical arcade games to cellular automaton simulations.
The crypto miners contain the predominant versions of the CryptoNight algorithm and their less memory and computation-intensive variants,.
Importantly, this provides extensive coverage of crypto mining malware, as several studies have found that in-the-wild cryptojacking implementations are all based on the CryptoNight algorithm~\cite{cabrera-arteaga-2022-webassemblydiversificationmalware, hilbig-2021-empiricalstudyreal}.

\subsection{Implementation}
\label{sec:implementation}

\begin{figure}[h]
	\begin{center}
		\includegraphics[width=1.0\textwidth]{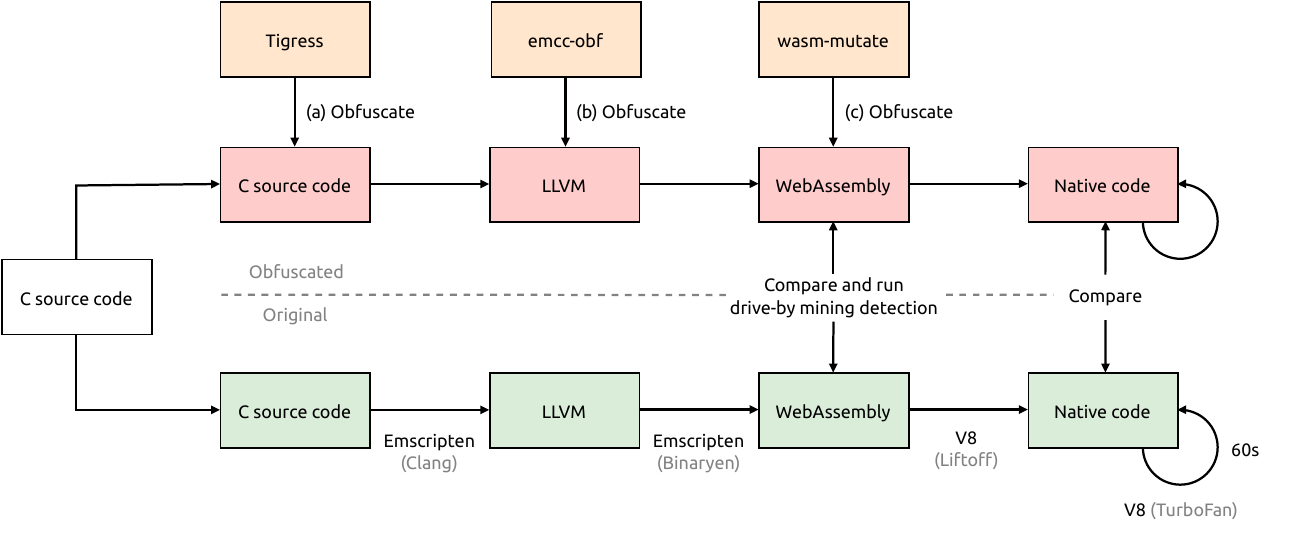}
	\end{center}
	\caption{Overview of the research strategy: Each program is obfuscated using either (a) Tigress, (b) emcc-obf, or (c) wasm-mutate at the corresponding  abstraction level, and finally compiled to WebAssembly. The WebAssembly binaries are then run through the cryptojacking detectors, instantiated in the browser to extract the native code, and compared with their non-obfuscated counterparts.}
	\label{fig:methodology-overview}
\end{figure}

The implementation, aligned with the research strategy depicted in Figure \ref{fig:methodology-overview}, is described in the following sections.
Each application in the dataset undergoes obfuscation using Tigress, emcc-obf, and wasm-mutate, as specified in Section \ref{sec:method-obfuscation}.
After obfuscation, the original and obfuscated binaries are fed to the cryptojacking detectors, detailed in Section \ref{sec:method-drive-by-mining}.
Then, the WebAssembly binaries are instantiated within the Chrome browser to extract the native code, as explained in Section \ref{sec:method-native}.
To measure the performance overhead of the crypto miners, we implement a cryptojacking client, web server, and WebSocket proxy server, as described in Section \ref{sec:method-measuring-hash-rate}.
Lastly, we implement the \gls{DTW} algorithm as a means of comparing the WebAssembly binaries, detailed in Section \ref{sec:method-dynamic-time-warping}.
In the interest of transparency and reproducibility, the code used to conduct the experiments is publicly available on GitHub.\footnote{\url{https://github.com/HakonHarnes/wasm-obf}}

\subsubsection{Obfuscation}
\label{sec:method-obfuscation}

For obfuscation, we use a variety of obfuscation methods, each operating at a different abstraction level.
We obfuscate the source code using Tigress, the LLVM bitcode using emcc-obf, and the WebAssembly code using wasm-mutate.
In other words; we either apply obfuscation \textit{before} compilation using Tigress, \textit{during} compilation using emcc-obf, or \textit{after} compilation using wasm-mutate.

The applications in Table \ref{tab:dataset} are obfuscated using Tigress, emcc-obf, and wasm-mutate, with the number of binaries generated for each application shown in Table \ref{tab:dataset-obfuscated}. 
For Tigress and emcc-obf, eight transformations are applied per application, while for wasm-mutate, six transformations are applied per application. 
However, since wasm-mutate produced unsatisfactory results when applying individual transformations, and in an effort to replicate results from other studies~\cite{cabrera-arteaga-2022-webassemblydiversificationmalware}, we also apply stacked transformations using wasm-mutate.
Stacked transformations involve applying one transformation after the other, in succession, to each application in the dataset. 
To this end, we apply a sequence of 1000 random transformations to each application
This process results in 22 WebAssembly binaries in the original dataset and a total of 22,484 binaries in the obfuscated dataset.

Since Tigress expects only one source file as input, we merge all the source files for each application into one file using the \gls{CIL}~\cite{necula2002cil}.
Although only Tigress requires a single source file, we use the same merged source file for emcc-obf and wasm-mutate to ensure that the source code is identical for all obfuscation methods.
To further ensure consistency, we compile all the WebAssembly binaries using the same Emscripten version, version 3.1.35.
The applications were compiled with optimization turned off to ensure the compiler did not optimize away the obfuscation applied.

To ensure the correctness of the obfuscated binaries, we invoke all of them in the browser and manually check that they are still functioning as intended. 
For the stacked wasm-mutate transformations, we perform this check every 100th iteration. In the case of crypto miners, we verify that the hashes reach the mining pool and are accepted by it as valid hashes. 
However, due to the mining network difficulty, we were only able to directly verify this for cn-r, cn-lite-0, and cn-pico-trtl, as the other variants could not solve and submit the hashes before receiving new jobs. 
To address this, we compare the first 100 hashes of the original and obfuscated binaries for all CryptoNight variants and ensure that they are identical. 
Through this process, we found that the applications in our dataset continued to function as intended after obfuscation.

\begin{table}[t]
	\small
	\begin{center}
		\begin{tabular}{llllll}
			\toprule
			Category      & Original & Tigress & emcc-obf & wasm-mutate & wasm-mutate (stacked) \\
			\midrule
			Utilities     & 6        & 48      & 48       & 36          & 6000                \\
			Games         & 6        & 48      & 48       & 36          & 6000               \\
			Crypto miners & 10       & 80      & 80       & 60          & 10,000              \\
			\midrule
			Sum           & 22       & 176     & 176      & 132         & 22,000              \\
			\bottomrule
		\end{tabular}
		\caption{Number of WebAssembly binaries in the original and obfuscated case.}
		\label{tab:dataset-obfuscated}
	\end{center}
\end{table}

\subsubsection{Tigress}

We use the latest version of Tigress, version 3.3.2, to obfuscate the source code of each application.

The following transformations were applied to the source code of each application:

\begin{itemize}
	\item \inlinesection{Flattening} Control obfuscation that transforms the control flow of an application into a flat hierarchy, thereby eliminating structured control flow.
	\item \inlinesection{Random Functions} Control obfuscation that generates a unique random function. Random function calls are also inserted into the generated code for increased complexity.
	\item \inlinesection{Function Splitting} Control obfuscation that splits a function into smaller sub-functions. This technique disguises the structure of the original function, thus complicating the process of code analysis.
	\item \inlinesection{Virtualization} Control obfuscation that transforms a function into a specialized interpreter by constructing a unique bytecode. This technique involves the creation of a virtual \gls{ISA} and a bytecode program, with each function essentially executing as a self-contained \gls{VM}.
	\item \inlinesection{Encode Arithmetic} Data obfuscation that replaces integer arithmetic with more complicated but equivalent expressions using \gls{MBA}~\cite{eyrolles2017obfuscation}.
	\item \inlinesection{Encode Literals} Data obfuscation that replaces constant integers and strings with code that dynamically generates them at runtime. Specifically, it uses opaque expressions to substitute integers and replaces strings with functions that generate them at runtime.
	\item \inlinesection{Anti-Alias Analysis} Preventive transformation that replaces all direct function calls with indirect ones to disrupt static analysis techniques that make use of alias analysis.
	\item \inlinesection{Anti-Taint Analysis} Preventive transformation that replaces the conventional data flow used for variable copying with control flow instead, with the aim of disrupting dynamic analysis tools that make use of taint analysis.
\end{itemize}

\subsubsection{emcc-obf}

We build and release emcc-obf,\footnote{\url{https://github.com/HakonHarnes/emcc-obf}} the first WebAssembly compiler with built-in obfuscation support.
Emcc-obf is a modified version of the Emscripten compiler, one of the most widely-used WebAssembly compilers.
The modifications are based on \gls{OLLVM}~\cite{junod-2015-ollvm}, which is no longer maintained.
Instead, we use the Hikari obfuscator,\footnote{\url{https://github.com/61bcdefg/Hikari-LLVM15}} a maintained fork of \gls{OLLVM} which is compatible with LLVM 16.0.0.
We build emcc-obf using the Hikari-modified version of LLVM 16.0.0, and compatible Binaryen and Emscripten versions.

The following transformations were applied to the LLVM-bitcode of each application:

\begin{itemize}
	\item \inlinesection{Control Flow Flattening} Control obfuscation that flattens the control flow of the program, similar to that of the flattening transformation of Tigress.
	\item \inlinesection{Bogus Control Flow} Control obfuscation that modifies the function call graph by inserting a new basic block preceding the original block. This new block includes an opaque predicate and executes a conditional jump to the original block.
	\item \inlinesection{Indirect Branches} Control obfuscation that replaces branching instructions with indirect branching. This technique thwarts disassemblers' ability to accurately predict the complete control flow through static analysis.
	\item \inlinesection{Basic Block Splitting} Control obfuscation that splits basic blocks, thereby breaking the structure by artificially increasing the number of basic blocks in a function.
	\item \inlinesection{Function Wrapper} Control obfuscation that encapsulates each target function within a generated wrapper function, introducing an additional layer of indirection to hinder control flow analysis.
	\item \inlinesection{Substitute Instruction} Data obfuscation that replaces arithmetic and boolean expressions with more complicated but equivalent instruction sequences. This is similar to the encode arithmetic transformation of Tigress.
	\item \inlinesection{Constants Encryption} Data obfuscation that encrypts constant integer values using the XOR cipher. The encrypted values are decrypted at runtime to their original form. This complicates reverse engineering as an analyst cannot directly read the constant values from the static code.
	\item \inlinesection{String Encryption} Data obfuscation that encrypts string values using the XOR cipher. The encrypted values are decrypted at runtime to their original form. Similar to constants encryption, but for strings.
\end{itemize}

\subsubsection{wasm-mutate}

We use the latest version ofwasm-tools,\footnote{\url{https://github.com/bytecodealliance/wasm-tools}} version 1.0.33, which contains the wasm-mutate tool.
We use the \texttt{--preserve\-semantics} flag to ensure that only semantics-preserving transformations are applied.

The following transformations were applied to WebAssembly code of each application:

\begin{itemize}
	\item \inlinesection{Code motion} Control obfuscation that modifies the \gls{AST} of a WebAssembly module by selectively applying a defined set of mutators, modifying the control flow or other aspects of the code while preserving its functionality.
	\item \inlinesection{Peephole} Data obfuscation that applies random, localized modifications on portions of the WebAssembly module. This is achieved by generating a minimal \gls{DFG} from a selected operator, applying predetermined rewriting rules to this DFG, resulting in a subtly modified version of the original segment in the module.
	\item \inlinesection{Add Function} Layout obfuscation that adds a function to the module.
	\item \inlinesection{Add Type} Layout obfuscation that adds a type to the module.
	\item \inlinesection{Add Custom Section} Layout obfuscation that adds a custom section to the module.
	\item \inlinesection{Remove Item} Layout obfuscation that removes an item (e.g. function) from the module.
\end{itemize}

\subsubsection{Cryptojacking Detection}
\label{sec:method-drive-by-mining}

In order to identify cryptojacking, we implement the detection methods presented in the background section; namely MINOS, WASim, MinerRay, and VirusTotal.
For MINOS, we use the reproduction by Cabrera et al.
We use the publicly available implementation for WASim, although we encountered challenges due to out-of-date dependencies, which we then updated to their latest versions.
We also use the public implementation of MinerRay, albeit faced with issues related to the JavaScript heap limit due to the path explosion problem. 
This led us to disable the function call linking for larger files as advised by the authors, although it may affect the detection rate.
Moreover, we implemented a two-minute timeout delay to prevent indefinite execution.
For VirusTotal, we use their API.

The MINOS reproduction by Cabrera et al., trained on 144 benign and 49 crypto mining binaries, resulted in a 0\% detection accuracy for our dataset.
To address this issue, we re-trained MINOS using the same dataset as Cabrera et al., augmenting it with the non-obfuscated binaries from our dataset.
We did not re-train the other machine learning-based detection methods as they delivered higher accuracies.

\subsubsection{Extracting the Native Code}
\label{sec:method-native}

Extracting the native code compiled by the V8 engine proved to be a challenging task.
After dialogue with the V8 developers, it was made clear that there is no convenient method to extract the native code generated by the V8 engine.
Despite this, we are able to determine the size of the native code that the V8 engine generates.
To do this, we instantiated the WebAssembly modules in the browser and let them run for 60 seconds, allowing time for TurboFan optimization.
Then, we used the \texttt{--print-wasm-code} flag in V8 to print the size of the native code generated by V8 for both Liftoff and TurboFan.
Although we could not directly extract the native code, its size provides insight into how, or if, obfuscation affects the native code.

\subsubsection{Measuring the Hash Rate}
\label{sec:method-measuring-hash-rate}

To measure the hash rate, we implement a cryptojacking client, web server, and WebSocket proxy server as described in Section \ref{sec:drive-by-mining}.
We use the public Webminerpool\footnote{\url{https://github.com/notgiven688/webminerpool}} implementation as a starting point, but we modify the code in several ways.
First, we extend the list of crypto mining pools so that we can support all CryptoNight variants.
Second, we fixed a bug in the code that caused the hash rate to be measured incorrectly.
Lastly, we containerized the client and server, as well as disabled caching, to ensure that the environment was consistent across all experiments.

\subsubsection{Dynamic Time Warping}
\label{sec:method-dynamic-time-warping}

We use \gls{DTW} to measure the dissimilarity between the WebAssembly binaries through the distance metric.
Since the \gls{DTW} algorithm expects numerical data, we preprocessed the WebAssembly binaries.
First, we convert the WebAssembly binaries from the binary format (\texttt{wasm}) to the human-readable format (\texttt{wat}).
Then, we use Python's built-in \texttt{hash} method to convert each instruction to a unique integer.
In practice, the WebAssembly binary is converted into a sequence of instructions represented as integers.
Given the considerable length of the WebAssembly binaries, we use FastDTW, an approximation of \gls{DTW} that reduces the overall time and space complexity~\cite{salvador2007toward}.

\subsection{Evaluation Metrics}
\label{sec:evaluation-metrics}

To address the research questions, we use several metrics to evaluate the effectiveness, detectability, and overhead introduced by obfuscation.
These metrics are presented in the following sections.

\subsubsection{RQ1 -- Effectiveness}

To evaluate obfuscation effectiveness, we use the distance between the original and obfuscated WebAssembly binaries, as derived from the \gls{DTW} algorithm.

\begin{definition}[Distance]
	\label{def:distance}
	The distance denotes the least cost of aligning two sequences of instructions, each representing a WebAssembly binary. The value represents the number of adjustments needed for one or both sequences to correspond to the other. As such, large distances indicate a substantial dissimilarity.
\end{definition}

Moreover, we investigate how obfuscation affects the size of the native code generated by the V8 engine and whether the TurboFan compiler can effectively eliminate instructions introduced by obfuscation.
To this end, we extract the native code size as described in Section \ref{sec:method-native} and compute the relative increase in native code size after obfuscation has been applied.
This is performed for the native code generated by both Liftoff and TurboFan.

\begin{definition}[Native code size increase]
	\label{def:native-code-size-increase}
	The increase in native code size indicates the relative increase in the size of the native code after obfuscation. This metric is calculated for the native code generated by the Liftoff and TurboFan compilers separately. 
\end{definition}

If $N$ denotes the size of the original native code, and $N'$ is the size after obfuscation, then:

\[
	\textrm{Native code size increase} = \frac{N' - N}{N}  \cdot 100
\]

\subsubsection{RQ2 -- Detectability}

To assess how effective the obfuscation methods are in evading detection, we feed the obfuscated binaries to the cryptojacking detectors presented in Section \ref{sec:method-drive-by-mining}.
Following this, we calculate the resulting precision, recall, and F$_1$ scores to assess the accuracy of the detection methods.
These measures are formally defined as follows:

\begin{definition}[Precision]
	\label{def:precision}
	Precision measures how many of the retrieved items are relevant. In the context of cryptojacking, it measures how many of the items identified as crypto miners are actually crypto miners.
\end{definition}
Precision is mathematically defined as:

\[
	\textrm{Precision} = \frac{\textrm{TP}}{\textrm{TP} + \textrm{FP}}
\]

\begin{definition}[Recall]
	\label{def:recall}
	Recall measures how many of the relevant items are retrieved. In the context of cryptojacking, it measures how many of the crypto miners are identified as crypto miners.
\end{definition}
Recall is mathematically defined as:
\[
	\textrm{Recall} = \frac{\textrm{TP}}{\textrm{TP} + \textrm{FN}}
\]
\begin{definition}
	[F$_1$ score]
	\label{def:f1-score}
	The F$_1$ score serves as a single metric that combines precision and recall. It is the harmonic mean of these two quantities, and hence, it gives equal weightage to both.
\end{definition}
The F$_1$ score is mathematically defined as:

\[
	\textrm{F}_1\ \textrm{score} = 2  \cdot \frac{\textrm{Precision}  \cdot \textrm{Recall}}{\textrm{Precision} + \textrm{Recall}}
\]

In this paper, we use the F$_1$ score as the primary metric to evaluate the overall accuracy of the detection methods.

\subsubsection{RQ3 -- Overhead}

We determine the size overhead by comparing the sizes of the original and obfuscated binaries.
The file size is measured in bytes using Python's \texttt{getsize} method. The relative increase in file size is then determined.

\begin{definition}[File size increase]
	\label{def:file-size-increase}
	The file size increase refers to the relative increase in file size caused by obfuscation. 
\end{definition}

If $S$ is the original file size and $S'$ the size after obfuscation, then:

\[
	\textrm{File size increase} = \frac{S' - S}{S} \cdot 100
\]

For the crypto mining binaries, we measure and compare the hash rates in the original and obfuscated cases.
To this end, we implement a cryptojacking setup as described in Section \ref{sec:method-drive-by-mining}.
We let the binaries calculate hashes for 100 seconds before measuring the total hashes.

\begin{definition}[Hash rate]
	\label{def:hash-rate}
	The hash rate is defined as the number of hashes calculated per second. 
\end{definition}

If $h$ is the total number of hashes calculated in time $t$, then:

\[
	\textrm{Hash rate} = \frac{h}{t}
\]

To quantify the performance overhead introduced by obfuscation, we calculate the relative hash rate of the obfuscated binaries compared to the original binaries.

\begin{definition}[Relative hash rate]
	\label{def:relative-hash-rate}
	The relative hash rate is a measure of the change in performance due to obfuscation. 
\end{definition}

If $H$ is the original hash rate, and $H'$ the hash rate after obfuscation, then:

\[
	\textrm{Relative hash rate} = \frac{H'}{H} \cdot 100
\]

% -------
% RESULTS
% -------

\section{Results} \label{ch:results}

In this section, we present the results of our experiments, addressing the research questions outlined in the introduction. We begin by examining the effectiveness of the obfuscation methods in producing dissimilar WebAssembly binaries and their impact on the resulting native code. We then assess the detectability of the obfuscated binaries by state-of-the-art cryptojacking detectors. Finally, we quantify the overhead introduced by the obfuscation methods in terms of file size and hash rate.

\subsection{Effectiveness}
\label{sec:res-effectiveness}

\subsubsection{Distances After Obfuscation}
\label{sec:res-effectiveness-distance}

\begin{figure}[ht]
	\centering
	\begin{subfigure}[T]{0.65\textwidth}
		\includegraphics[width=\textwidth]{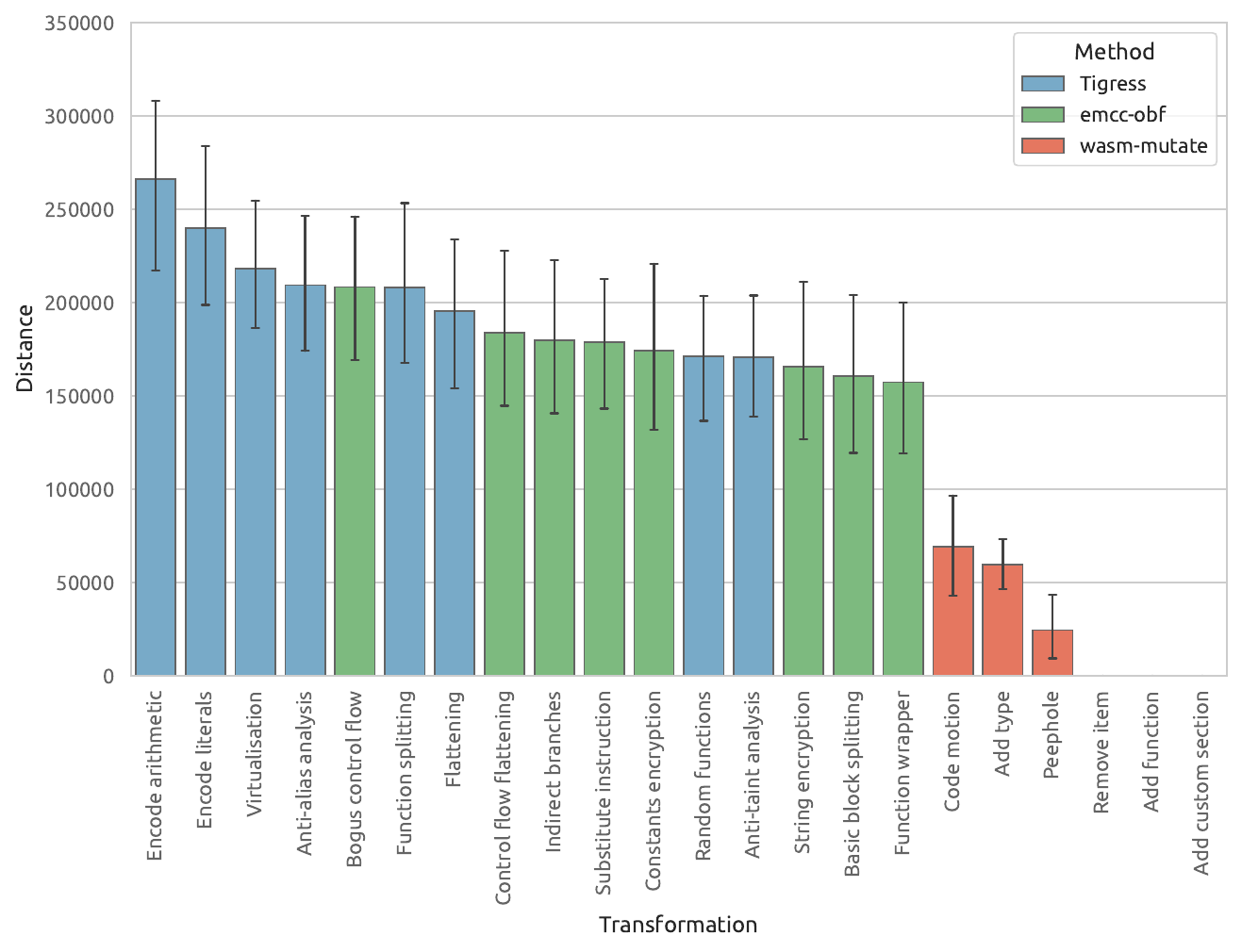}
		\caption{Distances for each transformation.}
		\label{fig:res-sub-distance-transformations}
	\end{subfigure}
	\hfill
	\begin{subfigure}[T]{0.338\textwidth}
		\includegraphics[width=\textwidth]{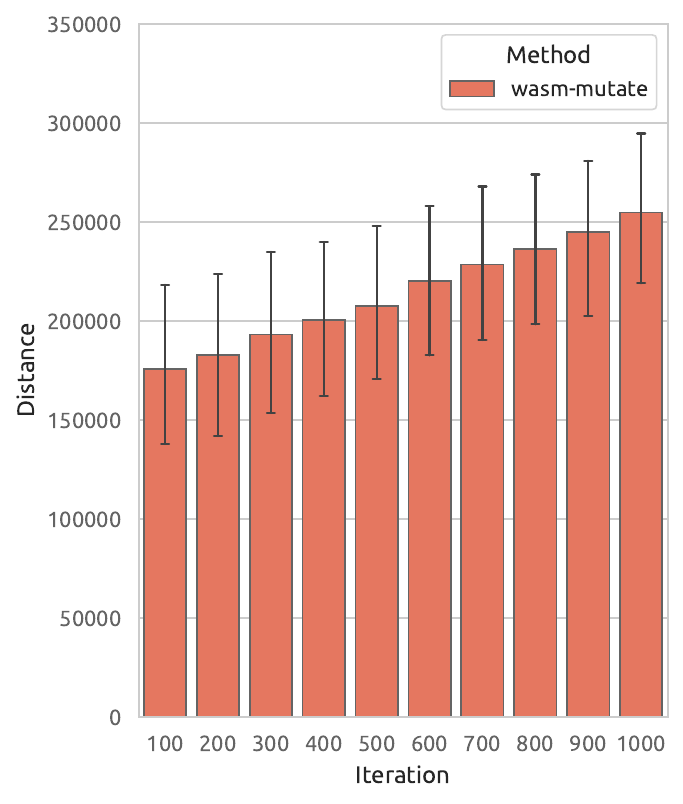}
		\vspace{3.15em}
		\caption{Distances for each iteration.}
		\label{fig:res-sub-distance-iteration}
	\end{subfigure}
	\caption{Distances for each obfuscation method, transformation, and iteration sorted in descending order. The error bars shown are indicative of a 95\% confidence interval.}
	\label{fig:res-effectiveness-distance}
\end{figure}

To assess the effectiveness of obfuscation methods, we measured the distances between the original and obfuscated WebAssembly binaries for each method, transformation, and iteration. The distance metric (Definition \ref{def:distance}) quantifies the dissimilarity between binaries, with larger distances indicating more effective obfuscation.

Figure \ref{fig:res-effectiveness-distance} presents the distances for each obfuscation method and transformation, revealing that Tigress is the most effective method with an average distance of 209,000, followed by emcc-obf (176,000) and wasm-mutate (30,000). Notably, wasm-mutate achieves distances up to 252,000 when transformations are stacked, rivaling Tigress in effectiveness. The most effective transformations overall are encode arithmetic, encode literals, and virtualization, all applied by Tigress. However, the type of obfuscation (data or control) that performs best varies by method. Data obfuscations like encode arithmetic and encode literals are more effective for Tigress, while control obfuscations such as control flow flattening and code motion are more effective for emcc-obf and wasm-mutate.

\begin{figure}[ht]
	\begin{center}
		\includegraphics[width=1.0\textwidth]{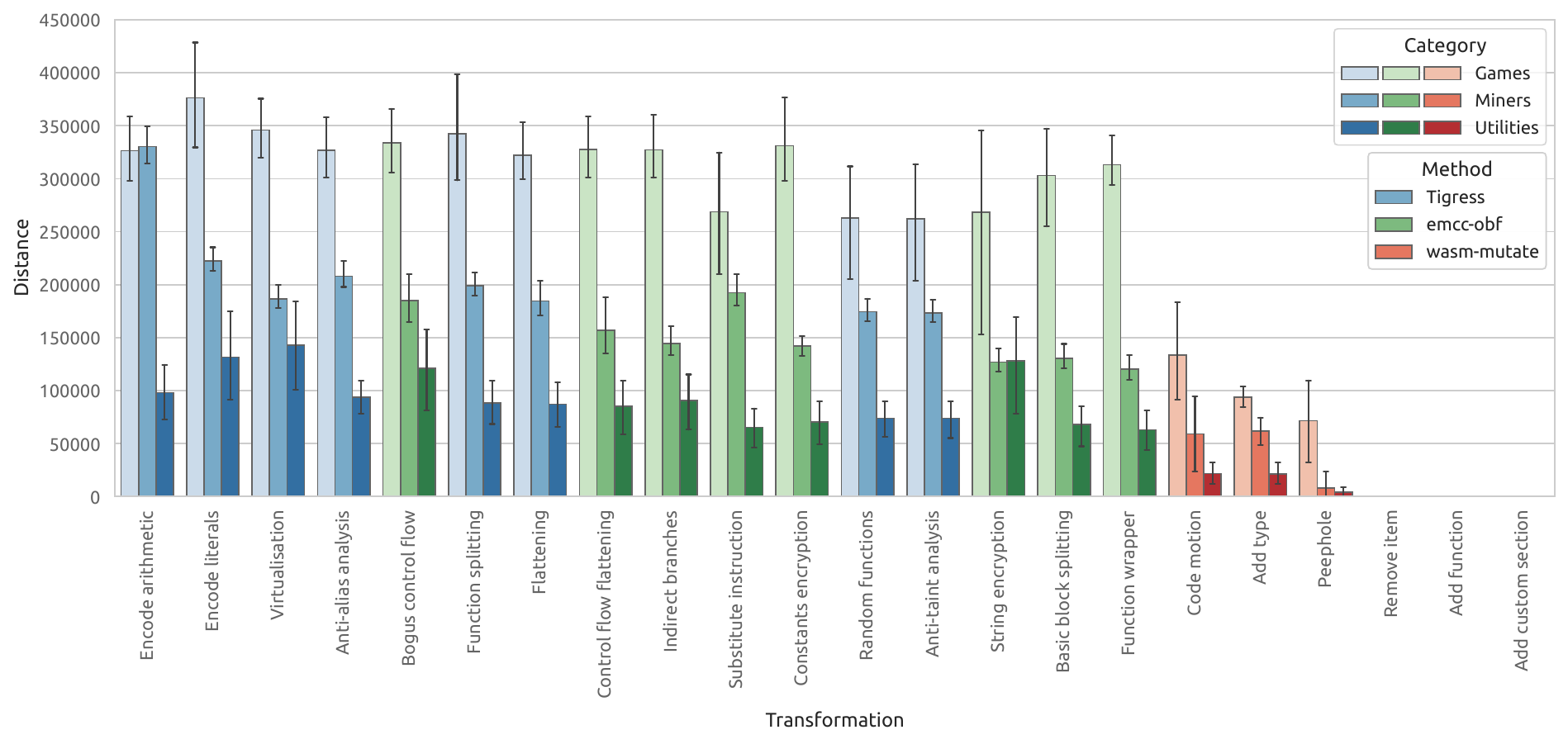}
	\end{center}
	\caption{Distances for each obfuscation method and transformation grouped by program category, sorted by the average distances in descending order. The error bars shown are indicative of a 95\% confidence interval.}
	\label{fig:res-effectiveness-distance-category}
\end{figure}

Figure \ref{fig:res-effectiveness-distance-category} shows the distances grouped by application category.
The noticeable variations in distances between application categories can be attributed to the average sizes of the applications within each category. Longer sequences usually lead to larger distances, and games are typically larger than utilities due to the inclusion of external libraries.
Therefore, the distances should not be compared directly across application categories.
Instead, the focus should be on the relative effectiveness of the transformations within each specific application category.

Comparing transformations within each category, we observe that the most effective transformations depend on the application type. 
For crypto miners, encode arithmetic and substitute instructions prove to be the most efficient. 
In the case of games, encode literals and constants encryption are most effective.
Conversely, virtualization and string encryption are most effective for utilities, although string encryption is the \textit{least} effective for games and crypto miners.
These observations highlight the fact that the effectiveness of the transformations is largely dependent on the nature of the application being obfuscated.

This observation is further reinforced by the observation that transformations found effective at one abstraction level also perform well at other abstraction levels.
Encode arithmetic (applied to the source code) and the similar substitute instructions (applied to the LLVM bitcode) are the most effective transformations for crypto miners.
Similarly, encode literals (applied to the source code) and the corresponding constants encryption (applied to the LLVM bitcode) are the most effective transformations for games.
This highlights the significant influence the content of the application has on the effectiveness of the transformations, regardless of the abstraction level at which transformations are applied.

\subsubsection{Native Code Size Increase}
\label{sec:res-effectiveness-native}

\begin{figure}[ht]
	\begin{center}
		\includegraphics[width=1.0\textwidth]{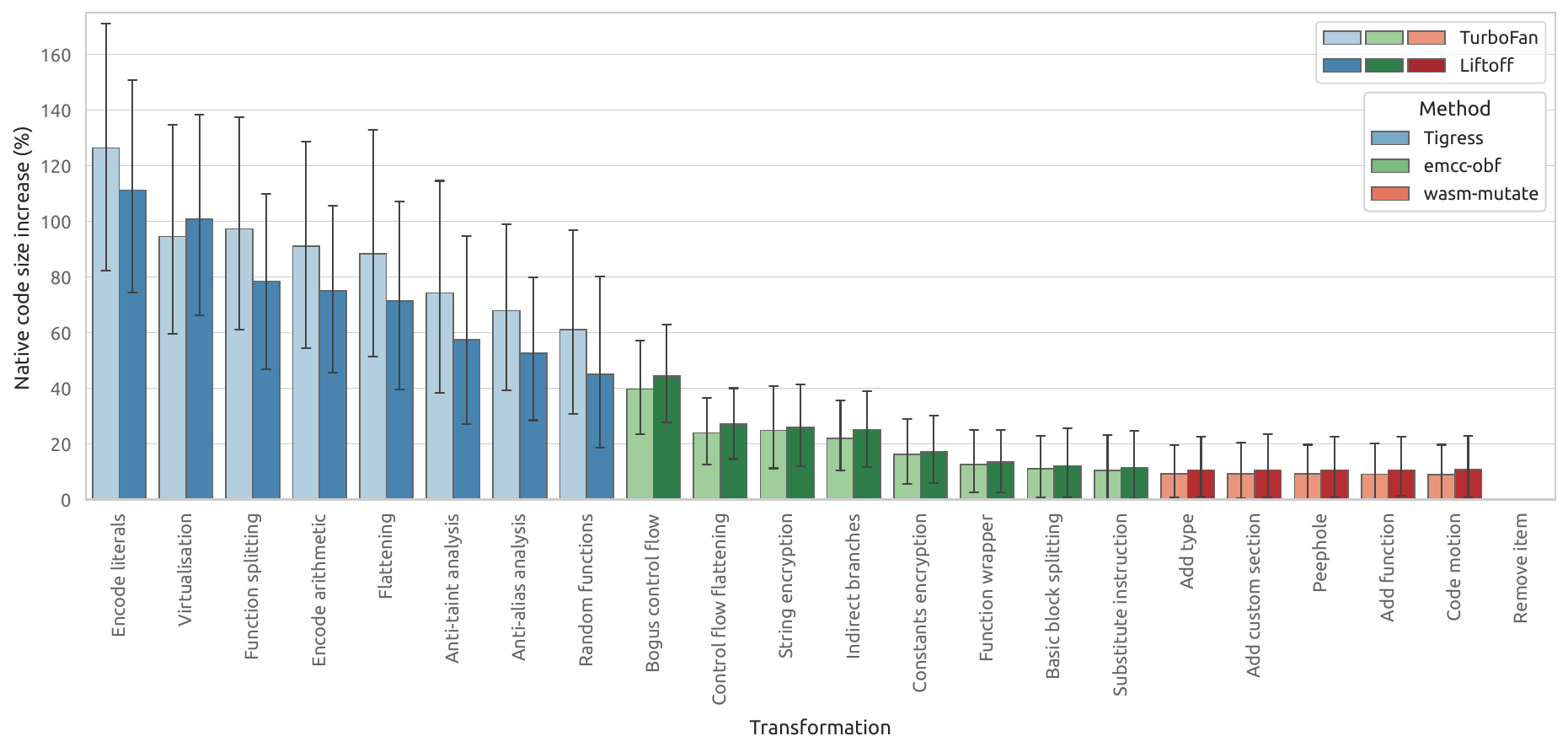}
	\end{center}
	\caption{Native code size increase for each obfuscation method and transformation after lazy compilation (Liftoff) and optimization (TurboFan) in the V8 engine, sorted by the average native code size increase in descending order. The error bars shown are indicative of a 95\% confidence interval.}
	\label{fig:res-effectiveness-distance-v8-transformations}
\end{figure}

To assess the impact of obfuscation on the size of native code, we measured the relative increase in native code size (Definition \ref{def:native-code-size-increase}) for each obfuscation method, transformation, and iteration after lazy compilation (Liftoff) and optimization (TurboFan) in the V8 engine. These values are calculated relative to the initial native code sizes before obfuscation for both Liftoff and TurboFan-produced code.

Figure \ref{fig:res-effectiveness-distance-v8-transformations} presents the relative increase in native code size for each obfuscation method and transformation.
Although TurboFan may show a larger relative increase in some instances, it still reduces the overall native code size by about 30\% on average compared to Liftoff.
For example, consider the situation where Liftoff initially generates 100MB of native code. 
TurboFan optimizes this to 50MB. 
After obfuscation, Liftoff's output increases to 200MB, and TurboFan's optimization reduces this to 150MB.
So, despite Liftoff showing a 100\% relative increase and TurboFan a 200\% relative increase after obfuscation, TurboFan's optimization still results in an overall reduction of native code in the original and obfuscated case.

\begin{figure}[t]
	\begin{center}
		\includegraphics[width=80mm]{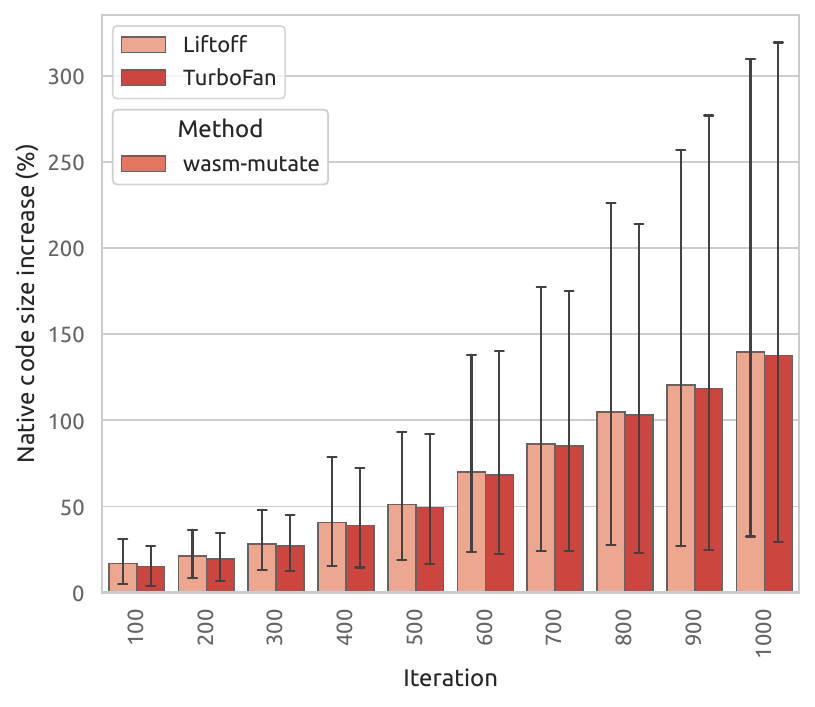}
	\end{center}
	\caption{Native code size increase for each iteration applied with wasm-mutate after lazy compilation (Liftoff) and after optimization (TurboFan) in the V8 engine. The error bars shown are indicative of a 95\% confidence interval.}
	\label{fig:res-effectiveness-distance-v8-mutate}
\end{figure}

Our findings show that the transformations from Tigress lead to the largest increase in native code, with an average of 87.25\% and 73.25\% after compilation by Liftoff and TurboFan, respectively.
Emcc-obf causes considerably smaller increases, averaging 20\% and 22\% after Liftoff and TurboFan compilation, while wasm-mutate demonstrates the slightest increase of 10\% for both Liftoff and TurboFan.
However, when stacking transformations, wasm-mutate substantially increases the native code size, with relative increases ranging from 16\% to 140\%, as shown in Figure \ref{fig:res-effectiveness-distance-v8-mutate}.

The impact of data and control obfuscations on native code size varies by obfuscation method. 
For Tigress, data obfuscations like encode literals cause a more substantial increase than control obfuscations like virtualization. 
In contrast, for emcc-obf, control obfuscations such as bogus control flow and control flow flattening impose a larger increase compared to data obfuscations like constants encryption and string encryption. 
For wasm-mutate, there are no discernible differences between control and data obfuscation.

Interestingly, for Tigress, the relative increase in native code is larger for Liftoff than for TurboFan, while the opposite is true for emcc-obf and wasm-mutate. 
In either case, TurboFan always decreased the size of the native code, but it is unlikely that it entirely eliminated the instructions introduced by obfuscation.

\subsubsection{Summary}
The effectiveness of WebAssembly obfuscation depends on the obfuscation method, transformation, and application being obfuscated. 
Tigress was found to be the most effective method, with encode arithmetic, encode literals, and virtualization being the most effective transformations overall. 
The type of obfuscation (data or control) that performs best varies by method, with data obfuscations being more effective for Tigress and control obfuscations being more effective for emcc-obf and wasm-mutate. 
The effectiveness of transformations also depends on the application type, with crypto miners benefiting from arithmetic encoding, games from literal encoding and constant encryption, and utilities from virtualization and string encryption. 
Obfuscation consistently increased the size of the native code generated by the V8 engine, with Tigress causing the largest increase. 
Although the V8 engine's TurboFan optimizer reduced the native code size by 30\% on average, it was unable to completely remove the additional instructions introduced by obfuscation.

\subsection{Detectability}
\label{sec:res-detectability}

\subsubsection{Detection Results}

\begin{figure}[ht]
	\centering
	\includegraphics[width=100mm]{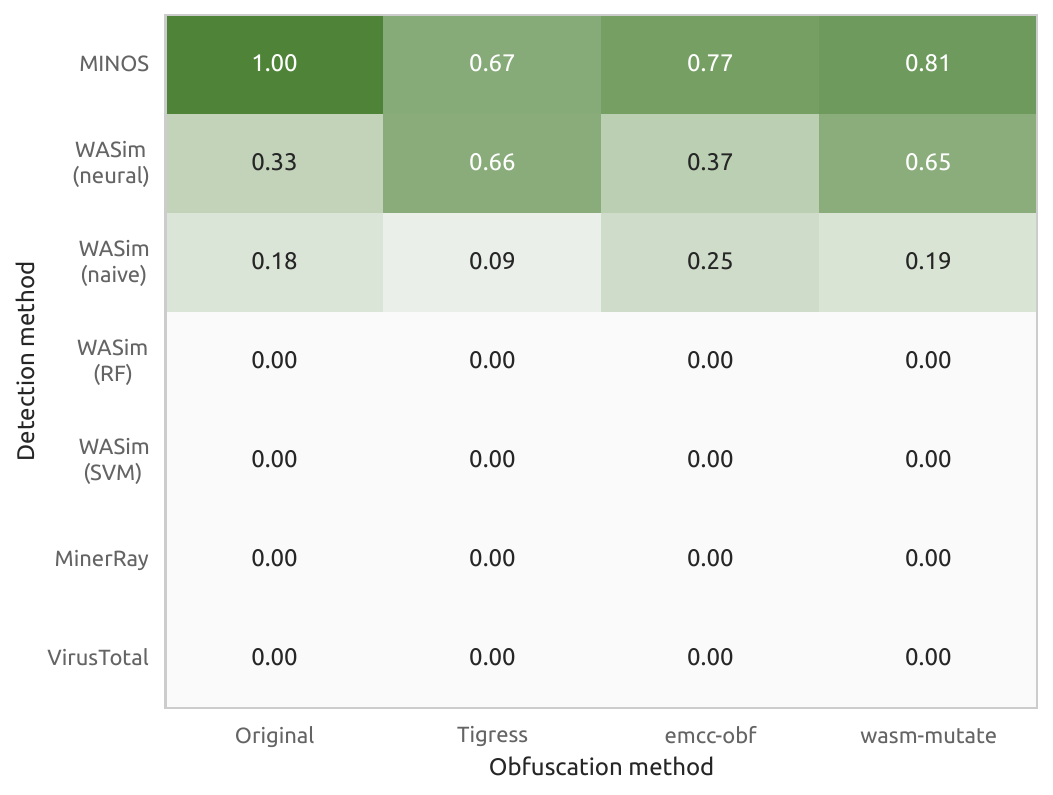}
	\caption{F$_1$ scores for each detection and obfuscation method. Darker colours indicate a higher F$_1$ score, while lighter colours indicate a lower F$_1$ score.}
	\label{fig:res-detectability-heatmap}
\end{figure}

To assess the effectiveness of the obfuscation methods in evading detection, we fed the obfuscated binaries to the cryptojacking detectors presented in Section \ref{sec:background-detecting-drive-by-mining}. 
We then calculated the resulting precision (Definition \ref{def:precision}), recall (Definition \ref{def:recall}), and F$_1$ scores (Definition \ref{def:f1-score}) to measure accuracy of the detection methods before and after obfuscation.

Figure \ref{fig:res-detectability-heatmap} shows the F$_1$ scores for each detection and obfuscation method.
We find that detection methods with higher accuracy tend to decrease in accuracy after obfuscation. 
For example, MINOS, having been re-trained for better accuracy, exhibits decreased accuracy after obfuscation. 
Here, Tigress proves the most effective, reducing the F$_1$ score from 1.0 to 0.67, with emcc-obf and wasm-mutate decreasing the F$_1$ score to 0.77 and 0.81, respectively.
In contrast, less accurate detection methods generally improve in accuracy after obfuscation. 
Both WASim (neural) and WASim (naive) see an increase in accuracy after obfuscation. 
Tigress, which was the most effective in decreasing the accuracy of MINOS, is the least effective for WASim (neural), increasing the F$_1$ score from 0.33 to 0.66. 
Similarly, emcc-obf and wasm-mutate also increase WASim accuracy (neural), albeit to a lesser extent, increasing the F$_1$ scores to 0.37 and 0.65, respectively.

Table \ref{tab:res-detection-minos-wasim} presents the precision, recall, and F$_1$ scores of MINOS and WASim (neural) after obfuscation. We exclude WASim (naive) and the other detection methods from the table due to their low accuracy.
From this, anti-alias analysis emerges as the most effective transformation, resulting in an F$_1$ score of 0 for both MINOS and WASim.
Several control obfuscations, such as flattening, control flow flattening, virtualization, and indirect branches, prove highly effective, resulting in F$_1$ scores of 0 for WASim.
However, not all control obfuscations achieve such results; function splitting and random functions increase the F$_1$ score of WASim to 0.95.
Despite data obfuscations like encode arithmetic and string encryption showing some effectiveness in evading detection, control obfuscation tends to be more effective overall.

\begin{table}[ht]
	\small
	\begin{tabularx}{\textwidth}{lXlXcccXccc}
		\toprule
		            &  &                         &  & \multicolumn{3}{c}{MINOS} &               & \multicolumn{3}{c}{WASim (neural)}                                                    \\
		\cmidrule{5-7} \cmidrule{9-11}
		Obfuscation &  & Transformation          &  & P*                        & R*            & F$_1$                              &  & P*            & R*            & F$_1$         \\
		\midrule
		Original    &  & None                    &  & 1.00                      & 1.00          & \textbf{1.00}                      &  & 1.00          & 0.20          & \textbf{0.33} \\
		\midrule
		Tigress     &  & Flattening              &  & 0.67                      & 1.00          & \textbf{0.80}                      &  & \textbf{0.00} & \textbf{0.00} & \textbf{0.00} \\
		            &  & Random functions        &  & 0.71                      & 1.00          & 0.83                               &  & 0.91          & 1.00          & \textbf{0.95} \\
		            &  & Function splitting      &  & 0.40                      & 0.20          & \textbf{0.27}                      &  & 0.91          & 1.00          & \textbf{0.95} \\
		            &  & Virtualization          &  & 0.75                      & 0.90          & 0.82                               &  & \textbf{0.00} & \textbf{0.00} & \textbf{0.00} \\
		            &  & Encode arithmetic       &  & 0.58                      & 0.70          & 0.63                               &  & 0.78          & 0.70          & 0.74          \\
		            &  & Encode literals         &  & 0.56                      & 1.00          & 0.72                               &  & 0.83          & 1.00          & 0.91          \\
		            &  & Anti-alias analysis     &  & \textbf{0.00}             & \textbf{0.00} & \textbf{0.00}                      &  & \textbf{0.00} & \textbf{0.00} & \textbf{0.00} \\
		            &  & Anti-taint analysis     &  & 0.77                      & 1.00          & 0.87                               &  & 0.89          & 0.80          & 0.84          \\
		\midrule
		emcc-obf    &  & Control flow flattening &  & 0.67                      & 1.00          & \textbf{0.80}                      &  & \textbf{0.00} & \textbf{0.00} & \textbf{0.00} \\
		            &  & Bogus control flow      &  & 0.55                      & 0.60          & 0.57                               &  & 0.60          & 0.30          & 0.40          \\
		            &  & Indirect branches       &  & 0.64                      & 0.90          & 0.75                               &  & \textbf{0.00} & \textbf{0.00} & \textbf{0.00} \\
		            &  & Basic block splitting   &  & 0.77                      & 1.00          & 0.87                               &  & 1.00          & 0.20          & 0.33          \\
		            &  & Function wrapper        &  & 0.91                      & 1.00          & 0.95                               &  & 1.00          & 0.60          & 0.75          \\
		            &  & Substitute instruction  &  & 0.73                      & 0.80          & 0.76                               &  & 1.00          & 0.50          & 0.67          \\
		            &  & Constants encryption    &  & 0.62                      & 0.80          & 0.70                               &  & 0.67          & 0.20          & 0.31          \\
		            &  & String encryption       &  & 0.62                      & 1.00          & 0.77                               &  & 0.40          & 0.20          & 0.27          \\
		\midrule
		wasm-mutate &  & Code motion             &  & 0.83                      & 1.00          & 0.91                               &  & 1.00          & 0.20          & 0.33          \\
		            &  & Peephole                &  & 0.91                      & 1.00          & 0.95                               &  & 1.00          & 0.20          & 0.33          \\
		            &  & Add function            &  & 0.91                      & 1.00          & 0.95                               &  & 1.00          & 0.20          & 0.33          \\
		            &  & Add Type                &  & 0.91                      & 1.00          & 0.95                               &  & 1.00          & 0.20          & 0.33          \\
		            &  & Add custom section      &  & 0.83                      & 1.00          & 0.91                               &  & 1.00          & 0.20          & 0.33          \\
		            &  & Remove item             &  & 0.91                      & 1.00          & 0.95                               &  & 1.00          & 0.20          & 0.33          \\
		\midrule
		wasm-mutate &  & Iteration 100           &  & 0.75                      & 0.90          & \textbf{0.82}                      &  & 0.75          & 0.30          & \textbf{0.43} \\
		            &  & Iteration 200           &  & 0.75                      & 0.90          & 0.82                               &  & 0.82          & 0.90          & 0.86          \\
		            &  & Iteration 300           &  & 0.69                      & 0.90          & 0.78                               &  & 0.88          & 0.70          & 0.78          \\
		            &  & Iteration 400           &  & 0.69                      & 0.90          & 0.78                               &  & 0.82          & 0.90          & 0.86          \\
		            &  & Iteration 500           &  & 0.69                      & 0.90          & \textbf{0.78}                      &  & 0.90          & 0.90          & \textbf{0.90} \\
		            &  & Iteration 600           &  & 0.67                      & 0.80          & 0.73                               &  & 1.00          & 0.80          & 0.89          \\
		            &  & Iteration 700           &  & 0.69                      & 0.90          & 0.78                               &  & 0.89          & 0.80          & 0.84          \\
		            &  & Iteration 800           &  & 0.67                      & 0.80          & 0.73                               &  & 0.88          & 0.70          & 0.78          \\
		            &  & Iteration 900           &  & 0.64                      & 0.70          & 0.67                               &  & 1.00          & 0.40          & 0.57          \\
		            &  & Iteration 1000          &  & 0.64                      & 0.70          & \textbf{0.67}                      &  & 1.00          & 0.20          & \textbf{0.33} \\
		\bottomrule
	\end{tabularx}
	\\[1mm]
	\raggedright
	\footnotesize{* Abbreviations: Precision (P), and recall (R).}
	\\[-1mm]
	\caption{Precision, recall, and F$_1$ scores for MINOS and WASim (neural) after applying obfuscation with Tigress, emcc-obf, and wasm-mutate.}
	\label{tab:res-detection-minos-wasim}
\end{table}

Moreover, we find that the effectiveness of the transformations varies significantly depending on the specific detection method.
While flattening is effective for WASim (F$_1$ score of 0), it is not nearly as effective for MINOS (F$_1$ score of 0.80).
Similarly, function splitting is effective for MINOS (F$_1$ score of 0.27) but not WASim (F$_1$ score of 0.95).

Interestingly, even after obfuscation, the recall often exceeds the precision for both MINOS and WASim , which suggests that the obfuscation methods are more effective at causing false positives than false negatives.
In other words, the drop in accuracy is primarily due to benign applications being mistakenly identified as crypto miners, rather than crypto miners evading detection.
There are certainly exceptions to this observation;
function splitting and bogus control flow effectively reduce recall for MINOS, and most control obfuscations do the same for WASim.

As observed previously, applying individual transformations with wasm-mutate does not significantly impact the accuracy of the detection methods.
However, stacking multiple transformations yields more promising results.
For MINOS, the F$_1$ score consistently decreases as more transformations are applied, indicating that the obfuscation becomes more effective at evading detection.
For WASim, however, the F$_1$ score inconsistently increases from 0.43 to 0.90 after 500 iterations, before falling back to 0.33 after 1000 iterations.
This suggests that the effectiveness of stacked transformations may not always scale linearly with the number of transformations applied, and there may be a point of diminishing returns or even a decrease in effectiveness.

\subsubsection{WASim Classifiers}

\begin{figure}[ht]
	\centering
	\hfill
	\begin{subfigure}[b]{0.4\textwidth}
		\includegraphics[width=\textwidth]{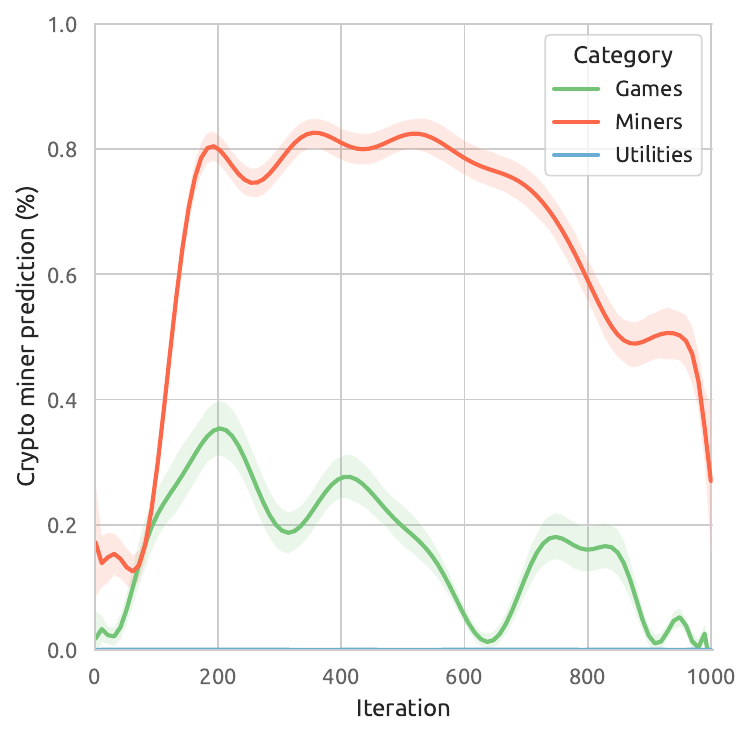}
		\caption{Neural.}
		\label{fig:res-sub-neural}
	\end{subfigure}
	\hfill
	\begin{subfigure}[b]{0.4\textwidth}
		\includegraphics[width=\textwidth]{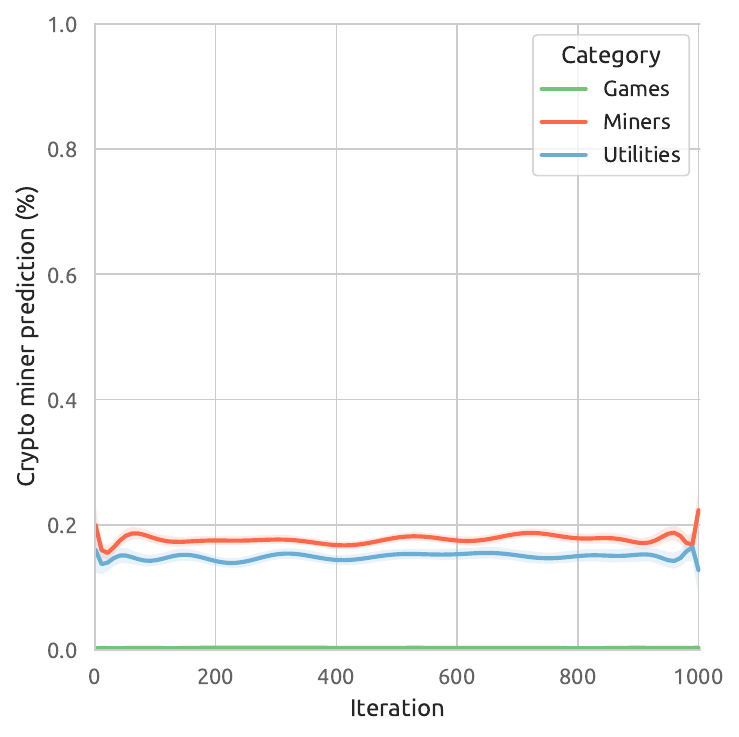}
		\caption{Naive Bayes.}
		\label{fig:res-sub-naive}
	\end{subfigure}
	\hfill
	\\[1em]
	\hfill
	\begin{subfigure}[b]{0.4\textwidth}
		\includegraphics[width=\textwidth]{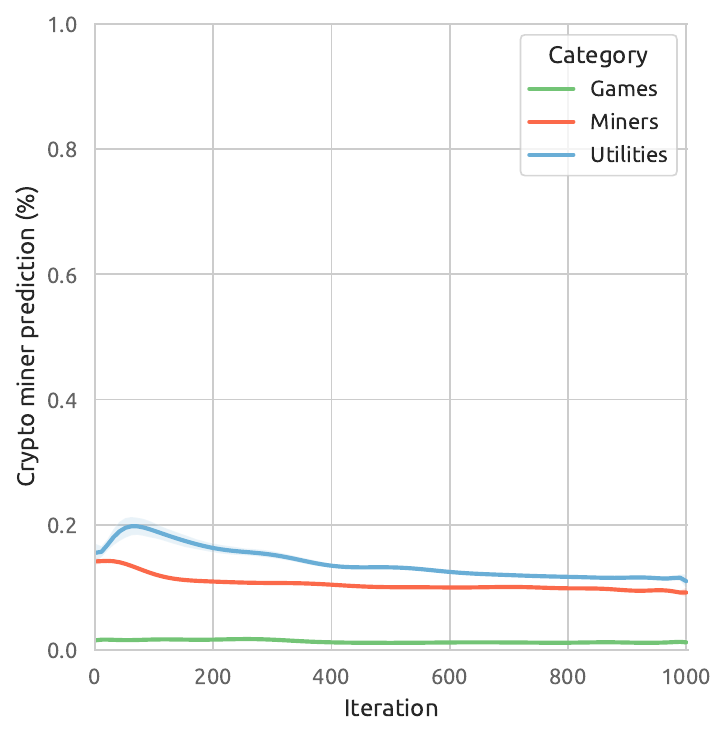}
		\caption{RF.}
		\label{fig:res-sub-random}
	\end{subfigure}
	\hfill
	\begin{subfigure}[b]{0.4\textwidth}
		\includegraphics[width=\textwidth]{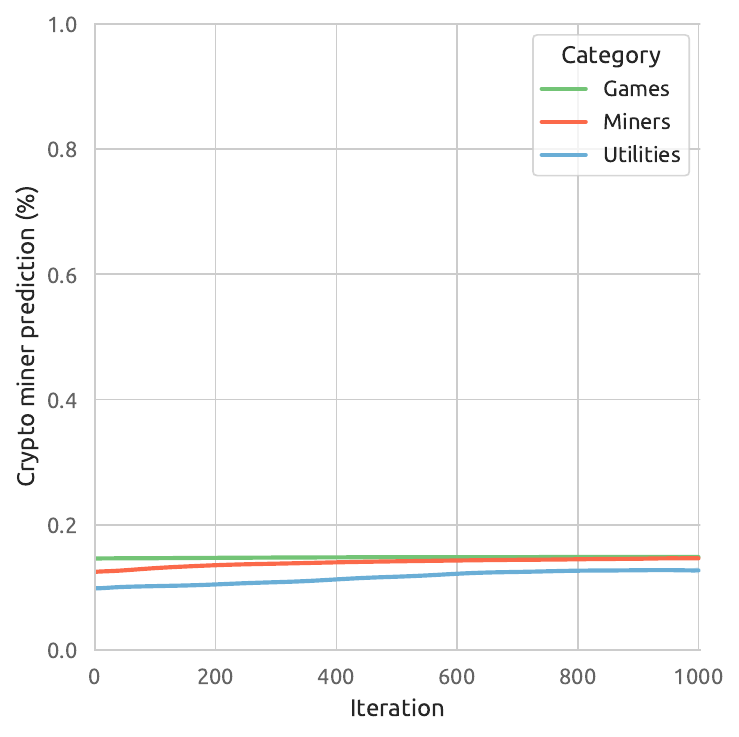}
		\caption{SVM.}
		\label{fig:res-sub-svm}
	\end{subfigure}
	\hfill
	\caption{The four different WASim classifiers and their respective prediction likelihoods for identifying a WebAssembly binary as a crypto miner as the binaries undergo iterative transformations using wasm-mutate. For each iteration, a randomly selected transformation is applied.}
	\label{fig:res-detectability-wasim-classifiers}
\end{figure}

Figure \ref{fig:res-detectability-wasim-classifiers} shows the predictions of the different WASim classifiers in response to stacked wasm-mutate transformations.
Predictions for the naive Bayes, RF, and \gls{SVM} classifiers remain relatively constant as more transformations are applied.
This shows that they are not affected by the transformations, indicating that they are resilient to obfuscation.
On the other hand, the neural classifier exhibits variance in its predictions as more transformations are applied, signifying that it is sensitive to the transformations.
Interestingly, this suggests that the neural classifier is not nearly as obfuscation-resilient as the other classifiers, despite being recommended by the authors of WASim.

\begin{figure}[ht]
	\centering
	\begin{subfigure}[T]{0.465\textwidth}
		\vspace{0.3em}
		\includegraphics[width=\textwidth]{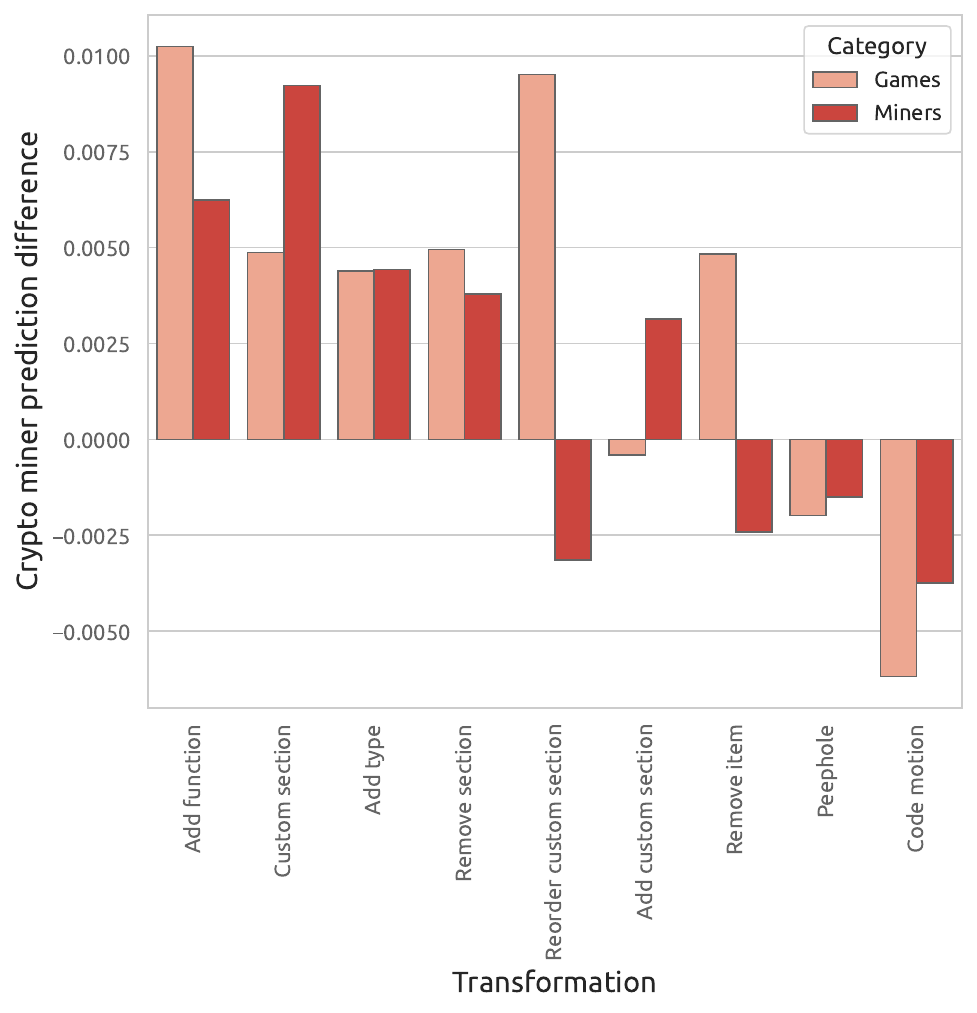}
		\caption{Most effective wasm-mutate transformations for altering the crypto miner predictions of the WASim (neural) classifier.}
		\label{fig:res-sub-mutation-effectiveness}
	\end{subfigure}
	\hfill
	\begin{subfigure}[T]{0.465\textwidth}
		\includegraphics[width=\textwidth]{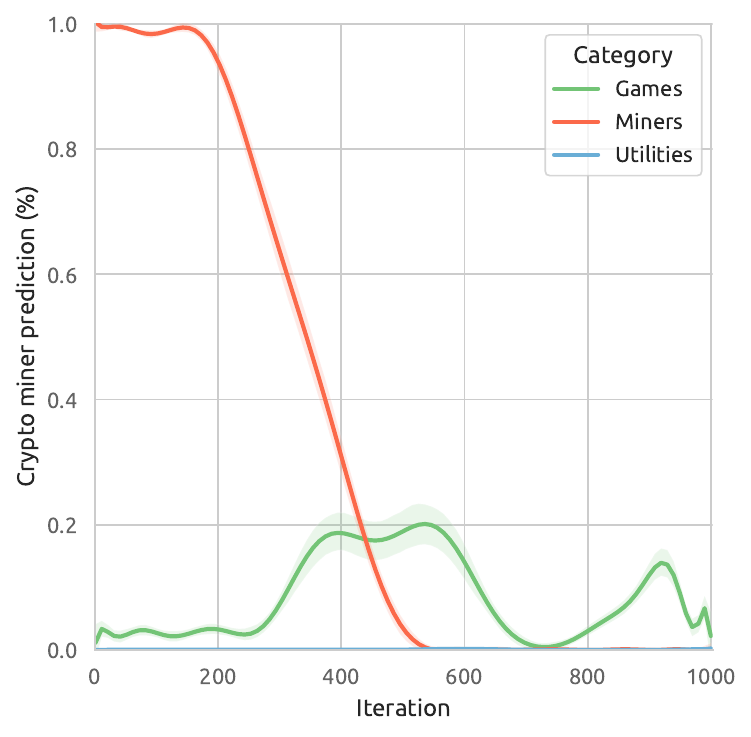}
		\vspace{0.1em}
		\caption{WASim (neural) predictions for WebAssembly binaries that have been iteratively obfuscated with the code motion and peephole transformations.}
		\label{fig:res-sub-wasim-scheme}
	\end{subfigure}
	\caption{Figure (a) shows the most effective wasm-mutate transformations for evading WASim (neural) detection. Figure (b) shows the predictions of WASim (neural) for WebAssembly binaries that have been iteratively obfuscated using the most effective transformations; namely code motion and peephole.}
	\label{fig:res-wasim-effective-and-scheme}
\end{figure}

This observation encouraged us to investigate which wasm-mutate transformations are the most effective for evading detection.
Specifically, which transformations decrease the crypto mining predictions of WASim (neural) the most?
As can be seen in Figure \ref{fig:res-sub-mutation-effectiveness}, the code motion and peephole transformations emerge as the most effective in reducing the crypto miner predictions, making them prime candidates for evading cryptojacking detection.

Intrigued by these findings, we explored the possibility of strategically applying these transformations to evade WASim (neural) detection.
First, we apply random mutations to the crypto mining binaries and select the resulting crypto mining binaries that have the highest crypto miner predictions.
Then, we iteratively apply the code motion and peephole transformations to those binaries and observe the predictions of the WASim (neural) classifier.
The results are shown in Figure \ref{fig:res-sub-wasim-scheme}.
The crypto miner binaries are initially labeled as crypto miners with 100\% probability.
Then, after 550 iterations, the crypto miner binaries are labeled as benign with 100\% probability, successfully evading detection and demonstrating how WASim (neural) can be strategically evaded using targeted transformations.

\subsubsection{Summary}
Obfuscation can be effective at evading state-of-the-art cryptojacking detectors, but its effectiveness depends on the specific detection method, obfuscation method, and transformation applied. 
Tigress proved most effective in evading MINOS, while emcc-obf was more effective against WASim. 
Anti-alias analysis was the only transformation that completely evaded both MINOS and WASim, but several control obfuscation transformations also achieved this for WASim. 
In general, control obfuscation was more effective than data obfuscation. 
Interestingly, the decrease in detection accuracy was often due to benign applications being misclassified as crypto miners, rather than crypto miners successfully evading detection. 
Stacking transformations with wasm-mutate yielded promising results, with the effectiveness increasing as more transformations were applied for MINOS, but fluctuating for WASim. 
However, by strategically applying the most effective transformations, namely code motion and peephole, wasm-mutate could completely evade WASim (neural) detection. 
The resilience of WASim's classifiers to obfuscation varied, with the neural classifier being more sensitive to transformations despite being recommended by WASim's authors.

\subsection{Overhead}
\label{sec:res-overhead}

\begin{figure}[ht]
	\centering
	\begin{subfigure}[T]{0.657\textwidth}
		\includegraphics[width=\textwidth]{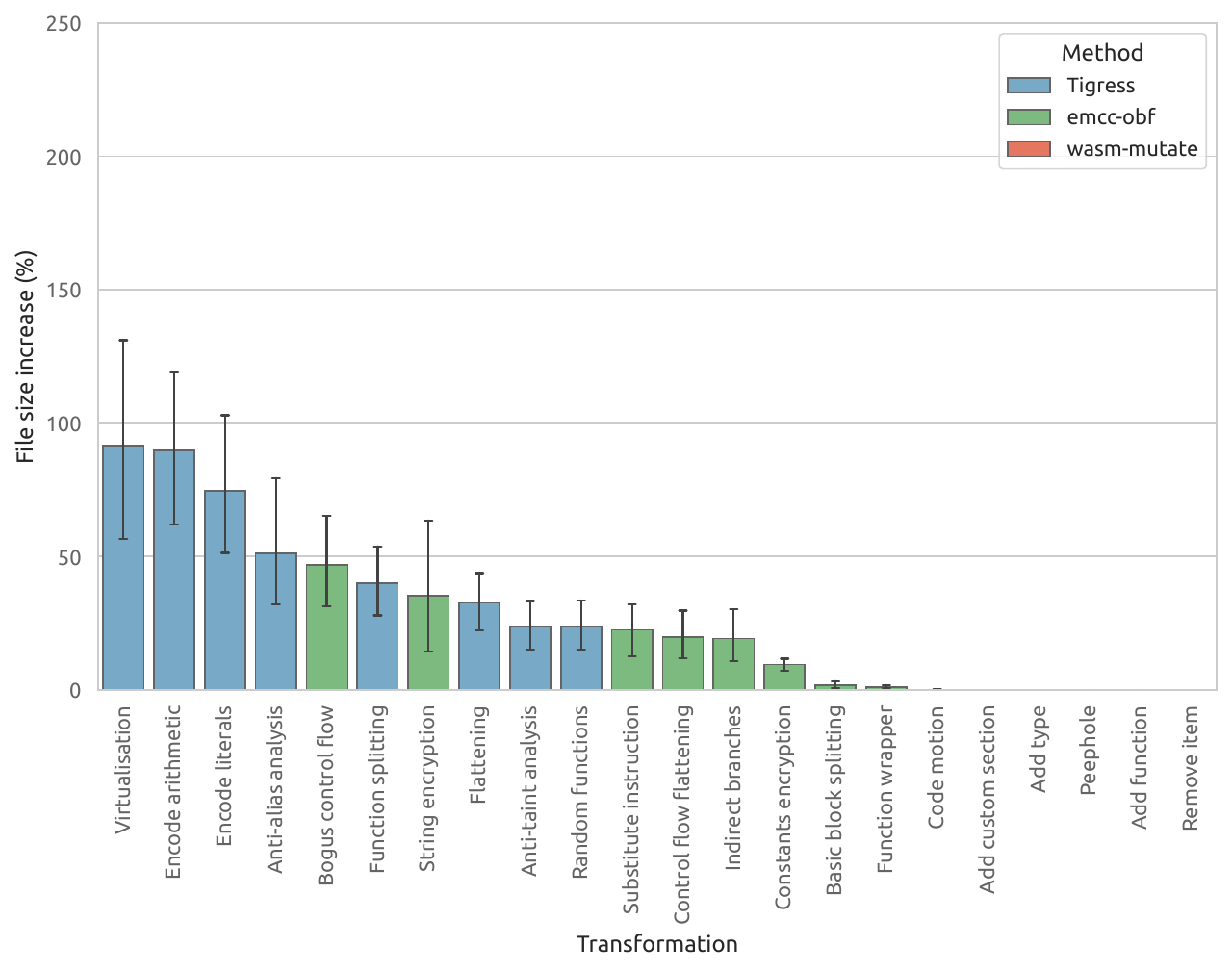}
		\caption{File sizes for each transformation.}
		\label{fig:res-sub-overhead-file-size-transformation}
	\end{subfigure}
	\hfill
	\begin{subfigure}[T]{0.332\textwidth}
		\includegraphics[width=\textwidth]{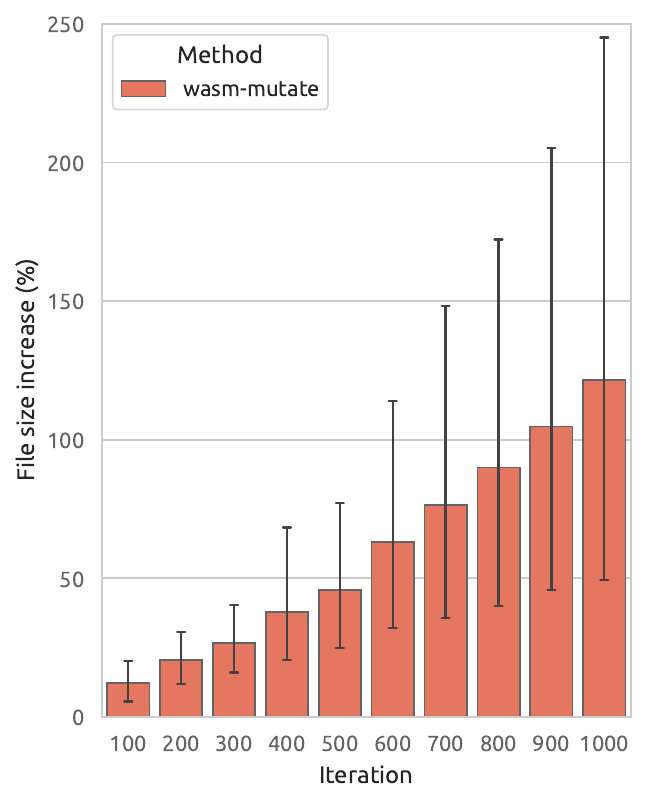}
		\vspace{3.35em}
		\caption{File sizes for each iteration.}
		\label{fig:res-sub-overhead-file-size-mutate}
	\end{subfigure}
	\caption{File size increase for each obfuscation method, transformation, and iteration sorted in descending order. The error bars shown are indicative of a 95\% confidence interval.}
	\label{fig:res-overhead-file-size}
\end{figure}

\subsubsection{File Size Overhead}

Figure \ref{fig:res-overhead-file-size} shows the relative increase in file size (Definition \ref{def:file-size-increase}) after obfuscation for each obfuscation method, transformation, and iteration.
Tigress increased the file size the most, averaging 53\%, followed by emcc-obf and wasm-mutate at 19\% and 0.2\%, respectively.
Despite wasm-mutate's minimal file size overhead when applying individual transformations, the overhead increases linearly when stacked transformations are applied, averaging a 59\% increase, ranging from 12\% to 121\%.
The transformations that contribute most significantly to file size increase are virtualization, encode arithmetic, and encode literals, which lead to increases of 91.6\%, 89.8\%, and 74.6\%, respectively.
Interestingly, no significant differences were observed between control and data obfuscation in terms of file size overhead.

\subsubsection{Hash Rate Overhead}

\begin{figure}[p]
	\centering
	\begin{subfigure}[T]{0.658\textwidth}
		\includegraphics[width=\textwidth]{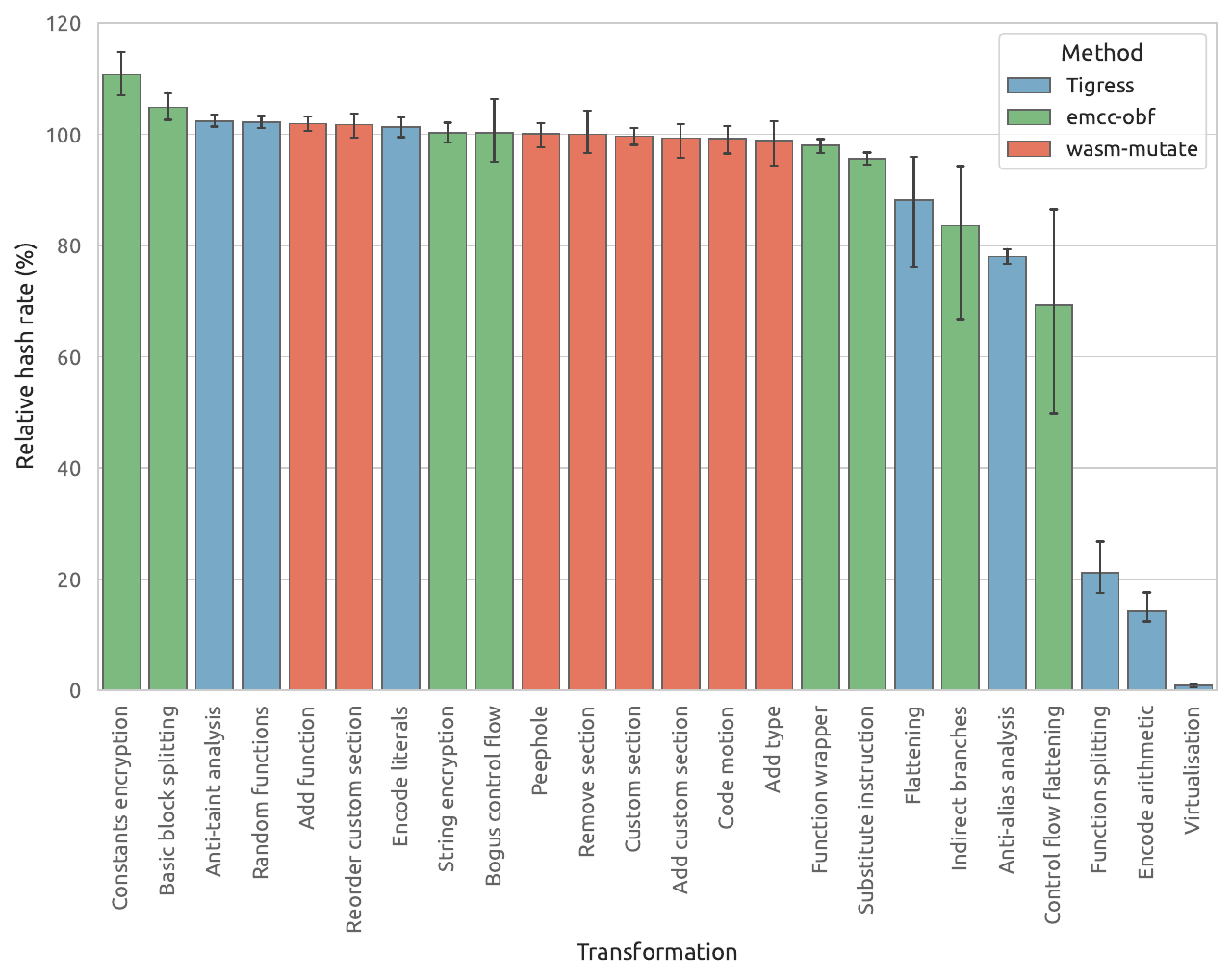}
		\caption{Hash rates for each transformation.}
		\label{fig:res-sub-overhead-hash-rate-transformations}
	\end{subfigure}
	\hfill
	\begin{subfigure}[T]{0.332\textwidth}
		\includegraphics[width=\textwidth]{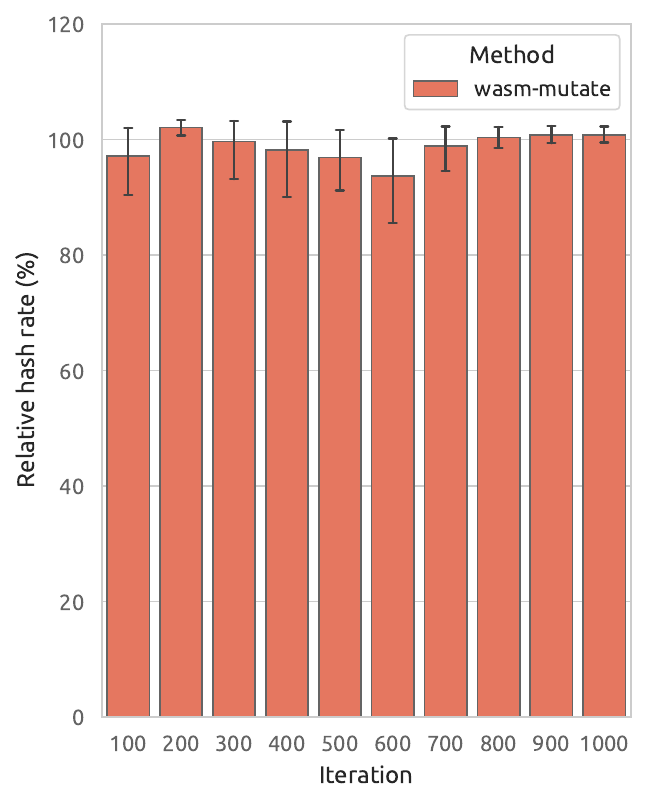}
		\vspace{3.4em}
		\caption{Hash rates for each iteration.}
		\label{fig:res-sub-overhead-hash-rate-mutate}
	\end{subfigure}
	\caption{Relative hash rates for each obfuscation method, transformation, and iteration sorted in descending order. The error bars shown are indicative of a 95\% confidence interval. }
	\label{fig:res-overhead-hash-rate}
	\vspace{2em}
	\begin{center}
		\includegraphics[width=\textwidth]{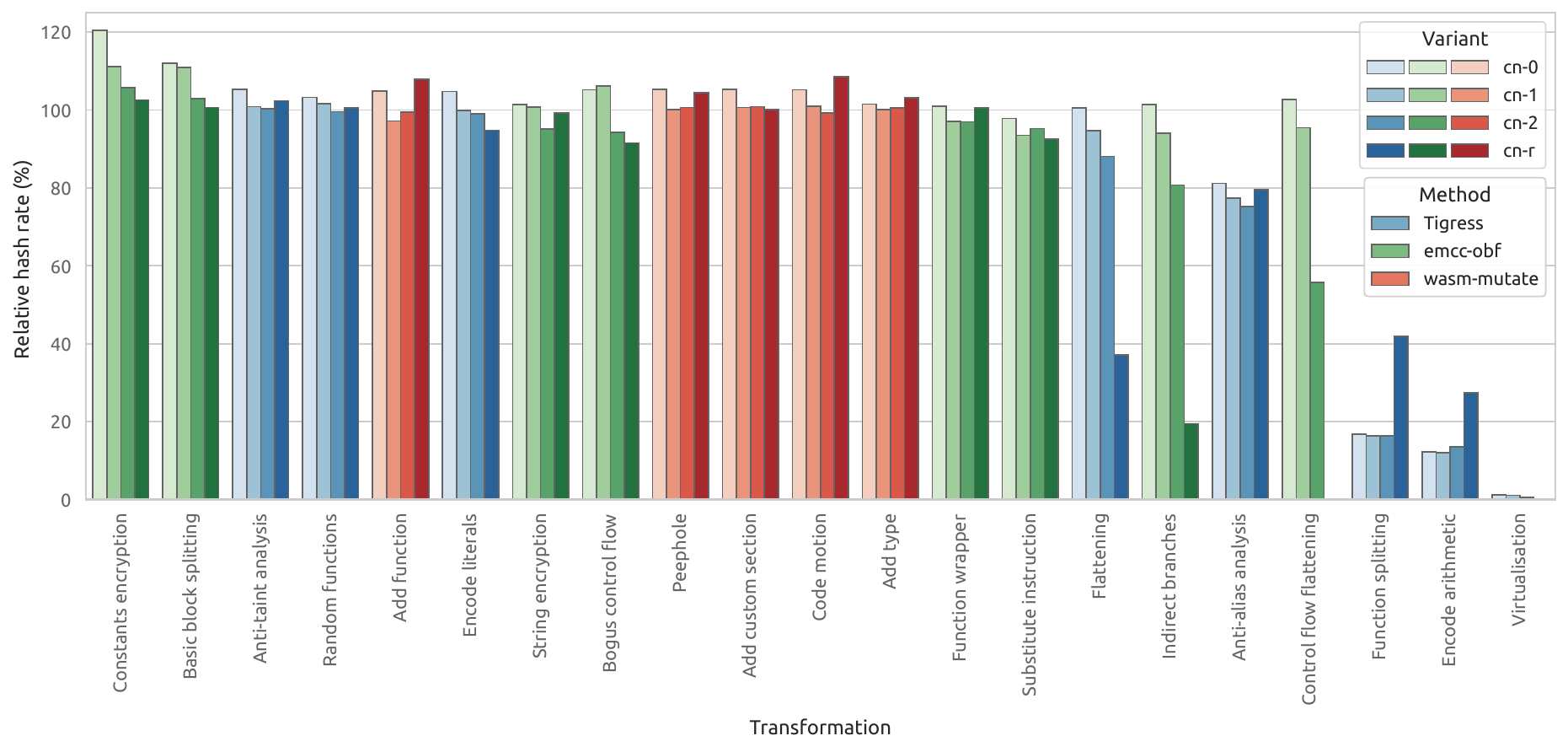}
	\end{center}
	\caption{Relative hash rates for each obfuscation method, transformation, and CryptoNight variant sorted in descending order.}
	\label{fig:res-overhead-hash-rate-variants}
\end{figure}

Figure \ref{fig:res-overhead-hash-rate} shows the relative hash rate (Definition \ref{def:relative-hash-rate}) for each obfuscation method, transformation, and iteration.
Tigress, emcc-obf, and wasm-mutate average 63.1\%, 95.2\%, and 99.8\% of the original hash rate, respectively.
Notably, two-thirds of the transformations do not significantly impact the hash rate.
Surprisingly, constant encryption and basic block splitting increase the hash rate by 110.7\% and 104.8\%, respectively.
Moderate overheads are seen in transformations like flattening, control flow flattening, indirect branches, and anti-alias analysis, delivering 70.2\% to 88.1\% of the original hash rate.
Significant overhead is introduced by Tigress' virtualization, encode arithmetic, and encode literals transformations, achieving 1\% to 21.1\% of the original hash rate.
Wasm-mutate imposes negligible hash rate overhead, which persists even with stacked transformations.
However, the hash rate fluctuates between iterations, ranging from 93.6\% to 100.8\% of the original hash rate, averaging 99.8\%.

Figure \ref{fig:res-overhead-hash-rate-variants} shows the relative hash rate (Definition \ref{def:relative-hash-rate}) after obfuscation for each obfuscation method, transformation, and CryptoNight variant.
There are noticeable differences between the variants.
Generally, cn-0 retains a higher hash rate compared to the other variants, indicating that it is less impacted by obfuscation.
There are also distinct differences between cn-r and the other variants.
Transformations like flattening, control flow flattening, and indirect branches bring about considerable overhead for cn-r but less so for the other variants.
Contrastingly, function splitting and encode arithmetic impose less overhead on cn-r but significantly impact the other variants.
This suggests that the transformation overhead is largely dependent on the specific CryptoNight variant being obfuscated.

\subsubsection{Summary}
Tigress consistently introduces the largest overheads, reflected in both the file size, averaging a 53\% increase, and the hash rate, averaging a reduction to 63.1\% of the original hash rate.
Moreover, Tigress' virtualization and encode arithmetic transformations consistently introduce the largest overheads for both file size and hash rate.
On the contrary, emcc-obf exhibits considerably less overhead, with a 19\% increase in file size and retention of 95.2\% of the original hash rate.
Remarkably, emcc-obf's constants encryption and basic block splitting transformations even increase the hash rate.
Although wasm-mutate introduces large file size overheads when stacking the transformations, ranging from 12\% to 121\%, it demonstrates minimal hash rate overhead, retaining 99.8\% of the original hash rate on average.
Additionally, variations between CryptoNight variants are observed, with cn-0 generally retaining a higher hash rate than other variants, while significant differences in hash rate are evident between cn-r and other variants, highlighting the importance of considering the specific variant when assessing the impact of obfuscation.

% ----------
% DISCUSSION 
% ----------

\section{Discussion} \label{ch:discussion}

In this section, we discuss the implications of our findings and provide insights into the effectiveness, detectability, and overhead of WebAssembly obfuscation. We analyze the factors that contribute to the varying effectiveness of the obfuscation methods and transformations, and we interpret the results in the context of real-world scenarios. We also discuss the limitations of our study and potential avenues for future research.

\subsection{Effectiveness}

Upon analyzing the WebAssembly binaries and resulting native code after obfuscation, we find that Tigress is generally the most effective method, followed by emcc-obf and wasm-mutate. 
This effectiveness is likely due to Tigress implementing more advanced transformations, as evidenced by the larger distances, increased binary size, and larger native code produced after compilation in the browser. 
These findings align with the research by Suresh et al.~\cite{suresh-2020-powerprofilinganalysis}, which concluded that Tigress generally applies more advanced transformations than OLLVM, the basis for emcc-obf.

We do not attribute the effectiveness of obfuscation methods to the abstraction level at which it is applied -- be it source code, LLVM, or WebAssembly -- but rather the implementation of the transformations themselves.
Implementing advanced transformations is generally simpler at higher abstraction levels due to the powerful abstractions provided by the programming language.
For example, implementing virtualization is simpler at the source code level, compared to implementing it in the low-level WebAssembly language.
However, lower abstraction levels allow for more targeted transformations, as seen with wasm-mutate's direct manipulation of the WebAssembly module.
Despite this, these targeted transformations are not advanced enough to be effective for obfuscation purposes when applied individually.

The ineffectiveness of individual transformations in wasm-mutate likely stems from its design as a diversifier rather than an obfuscator. 
Wasm-mutate makes subtle modifications to the WebAssembly module, allowing it to produce many diversified variants. 
This is in contrast to Tigress and emcc-obf, which implement advanced transformations that significantly modify the program. 
However, the results suggest that stacking wasm-mutate transformations can be an effective approach, which aligns with previous studies~\cite{cabrera-arteaga-2022-webassemblydiversificationmalware}.

Our findings indicate that the most effective transformations are encode arithmetic, encode literals, and virtualization.
In contrast, Bhansali et al.~\cite{bhansali-2022-firstlookcode} found anti-alias analysis to be the most effective, followed by virtualization and flattening.
We both find that the application's content significantly impacts transformation effectiveness.
In our findings, crypto miners benefit the most from the encode arithmetic transformation, likely due to the large number of arithmetic operations required for calculating the hashes.
Games benefit most from encode literals due to the large number of integer values used for scores, health points, and colours, to name a few.
Lastly, utilities benefit most from string encryption due to the large number of strings used for printing to the console and interacting with the user.
Contrastingly, Bhansali et al. identified anti-alias analysis as the most effective for all application types, except crypto miners, where virtualization was deemed most effective.
These differences likely stem from our divergent comparison methods; Bhansali et al. used cosine similarity, while we used a distance-based metric more suitable for WebAssembly.

The error bars in the plots further highlight the crucial influence of the application content on the effectiveness of the transformations.
For instance, Figure \ref{fig:res-effectiveness-distance-category} has a significant error bar for string encryption within the games category, implying that the effectiveness of this transformation depends on the amount of text used in each game.
As text content in games can significantly vary, the impact of string encryption will differ accordingly.

It is also imperative to highlight that the significant error bars, such as those observed in Figure \ref{fig:res-effectiveness-distance-v8-mutate}, stem from the heterogeneity of the dataset and the consequent variation in application sizes.
Games, for instance, tend to be larger than utilities due to the incorporation of external libraries.
Longer sequences tend to produce larger distances, accounting for the substantial error bars in these plots.

Tigress consistently produced more native code than emcc-obf and wasm-mutate, but this increase is not directly reflected in the WebAssembly binary size. 
We hypothesize that this discrepancy is due to Liftoff's lazy compilation strategy, which only compiles invoked functions.
Emcc-obf likely generates code that is never executed and thus never compiled to native code, resulting in large WebAssembly binaries but smaller native code, while Tigress generates code that is executed and compiled to native code.

TurboFan is more efficient at optimizing native code obfuscated using Tigress than with emcc-obf or wasm-mutate, likely because Tigress-produced native code contains more ``junk'' code that is easier to optimize away. 
Although TurboFan decreased the native code size by an average of 30\%, it is doubtful that it entirely eliminated the instructions introduced by obfuscation. 
Even for individual wasm-mutate transformations, TurboFan was unable to completely remove the extraneous instructions. 
Considering the significant increase in native code even after optimization, it is likely that at least some obfuscation remained. 
These findings contrast with previous studies on LLVM diversification for WebAssembly, where the native code was found to be identical after TurboFan optimization~\cite{arteaga-2020-crowcodediversification}.

\subsection{Detectability}

The experiments demonstrate that obfuscation techniques can successfully evade state-of-the-art cryptojacking detectors. 
For MINOS, this is not surprising, as obfuscation significantly alters the WebAssembly binaries, which are converted into grayscale images and used as input for the classifier. 
The obfuscation process likely leads to misclassification of the obfuscated program because the resulting images deviate considerably from the ones used during the training of the neural network. 
This hypothesis is supported by the findings of Loose et al.~\cite{Loose2023}, who demonstrated that modifying the grayscale image input of MINOS could lead to misclassification.

WASim takes a feature-based approach, analyzing properties like the binary size. 
As more transformations are applied (up to iteration 500), the crypto miner probability gradually increases, potentially because the classifier has learned that miners typically have larger binary sizes than other applications. 
However, as the binary size continues increasing beyond iteration 500, the probability decreases, possibly because the program becomes too large and complex to fit the expected profile of a miner. 
WASim's neural classifier proves particularly vulnerable to obfuscation, consistent with the known sensitivity of neural networks to adversarial inputs~\cite{szegedy-2013-intriguingpropertiesneural, goodfellow-2014-explainingharnessingadversarial, Loose2023}.

Notably, decreased detection rates often stemmed from benign applications being misclassified as miners rather than miners evading detection.
This is likely attributed to obfuscation increasing program complexity, leading to an over-classification of benign programs labeled as malicious. 
Similar findings were reported by Bhansali et al.~\cite{bhansali-2022-firstlookcode}, who found that obfuscation increased the false positive rate to 70\%.
It is important to stress that in real-world scenarios, false positives are detrimental, as they could result in harmless programs being unnecessarily blocked, degrading the user experience.
Since obfuscation can have legitimate uses, like safeguarding intellectual property, it is crucial to minimize its potential for inducing false positives.

In terms of transformations that were effective in evading detection, we generally observed that control obfuscation was more effective than data obfuscation.
This can be attributed to the fact that control obfuscation generally impacts the entire program, while data obfuscation targets specific parts, that is, the data.
As such, control obfuscation often results in more significant changes to the program, which is more likely to evade detection.
Intriguingly, anti-alias analysis, a preventive transformation, was the most effective for malware evasion.
Although neither MINOS nor WASim directly performs alias analysis, it was the only effective transformation for both detection methods.
Evidently, anti-alias analysis significantly impacts the features that the neural networks rely on, leading to misclassification.
On the contrary, Bhansali et al. found every Tigress transformations to be effective in evading MINOS~\cite{bhansali-2022-firstlookcode}.
However, they used the original MINOS implementation, while we used a reproduction by Cabrera et al. \cite{cabrera-arteaga-2022-webassemblydiversificationmalware}, which could explain the difference in results.
In addition, they used a small dataset of only two crypto miners, which threatens the validity of their results.

Wasm-mutate was found to be ineffective at evading detection when applying individual transformations. 
However, when the transformations were stacked, the accuracy of the detection methods significantly decreased.
We found that code motion and peephole were the most effective transformations.
Similarly, Cabrera et al.~\cite{cabrera-arteaga-2022-webassemblydiversificationmalware} found that peephole, add function, and code motion were the most effective transformations for evading detection.
They also found that it took between 120 and 635 iterations to evade detection, akin to our average of 550 iterations.
However, Cabrera et al. could completely evade MINOS detection using random transformations in under 1000 iterations.
We were unable to reproduce this, as we were only able to reduce the F$_1$ score to 0.67 after 1000 iterations.
This difference, we believe, arises from using different datasets.

\subsection{Overhead}

We found that Tigress and emcc-obf, on average, increased the WebAssembly binary size by 53\% and 19\%, respectively.
These observations align with Suresh et al.'s research~\cite{suresh-2020-powerprofilinganalysis}, wherein the code generated by Tigress was 5\% to 78\% larger compared to the code generated by OLLVM.
Notably, even for similar transformations like control flow flattening, Tigress produces 11.3\% larger binaries than emcc-obf. 
We attribute this to Tigress likely using a more complex control flow flattening algorithm.

While prior work did not measure wasm-mutate's impact on binary size ~\cite{cabreraarteaga-2022-artificialsoftwarediversification}, we found that stacking transformations can induce considerable overhead (12\% to 120\%, averaging 59\%). 
However, strategically selecting transformations to minimize overhead or applying fewer transformations can mitigate this issue, as demonstrated in other domains like Rosette language diversification~\cite{lundquist2016searching}.

In terms of performance overhead, Tigress performed worst (averaging 63.1\% of original hash rate), followed by emcc-obf (95.2\%) and wasm-mutate (99.8\%). 
These findings align with prior observations of Tigress-generated code causing a 4\% to 55.4\% larger performance overhead compared to OLLVM-generated code~\cite{suresh-2020-powerprofilinganalysis}. 
Function splitting, encode arithmetic, and virtualization induced the largest hash rate overheads and native code size increases. 
This correlation is unsurprising, as more native code means more instructions for the CPU to execute, reducing performance and hash rate. 
Tigress's advanced transformations significantly increase native code generation, resulting in lower hash rates.

Interestingly, despite the considerable native code increase from stacking 1000 wasm-mutate transformations, the performance overhead was negligible. 
We attribute this to hardware optimizations (e.g., pipelining, instruction reordering, branch prediction, cache optimization) applied to the native code, streamlining execution. 
Tigress's advanced transformations likely generate code too complex for the same degree of hardware optimization, resulting in larger performance overheads even with similar native code increases.

Another noteworthy observation is that constants encryption and basic block splitting increase the hash rate to 110.7\% and 104.8\% of the original hash rate, respectively.
This may be due to emcc-obf's clean-up passes designed to remove the intermediate values used for obfuscation.
We hypothesize that these clean-up procedures inadvertently optimize the code, leading to increased performance.

The WebAssembly binaries that were obfuscated with wasm-mutate fluctuated around the original hash rate, ranging from 93.6\% to 100.8\% of the original hash rate.
We attribute this to wasm-mutate performing transformations in the executable code of the module, which effectively work as optimizations.
For example, loop unrolling and code replacements can lead to smaller binaries and inadvertently increase the hash rate.
Conversly, certain transformations applied by wasm-mutate may have a detrimental effect on performance, accounting for the observed reductions in hash rate.
This hypothesis aligns with the findings of Cabrera et al.~\cite{cabreraarteaga-2022-artificialsoftwarediversification}, who reported wide fluctuations in performance overhead caused by wasm-mutate.

We discovered differences in performance overhead among CryptoNight variants. 
CryptoNightR (cn-r) is designed to be more resistant to \gls{ASIC} than its counterparts.
This is accomplished by embedding a random component into the algorithm, requiring miners to perform different operations for each block.
As a result, CryptoNightR uses a combination of arithmetic and branching operations, with the sequence randomized for each block.
Some of these arithmetic operations are generated at runtime, meaning these operations cannot be statically encoded, which explains why encode arithmetic is ineffective for CryptoNightR.
Additionally, CryptoNightR has 1.7 times more branch instructions than other variants, explaining the larger impact of flattening, control flow flattening, and indirect branches.

The increasing complexity of CryptoNight algorithms to ensure ASIC-resistance correlates with greater performance overheads from obfuscation. 
More complex algorithms with more operations and intricate control flows provide more "surface area" for obfuscation to slow down the program. 
The results show that cn-0 is usually least affected by obfuscation, succeeded by cn-1, cn-2, and cn-r, supporting this hypothesis.

\subsection{Interpreting the Findings}

The results derived from this paper are intricate and influenced by many factors.
Although several transformations can evade cryptojacking detection, they often introduce significant overhead.
This raises the question of whether obfuscation can be used for evading cryptojacking detection in real-world scenarios with justifiable overhead.
Our assessment suggests that it is feasible, but it depends on the obfuscation and detection methods, as well as the specific crypto miner algorithm.

For instance, anti-alias analysis can evade detection for both MINOS and WASim with every CryptoNight variant, albeit with a 22\% reduction in the original hash rate and a 51\% increase in file size. 
However, more desirable results can be achieved by adapting the obfuscation strategy to the specific use case. 
As an example, the original CryptoNight algorithm can be obfuscated using indirect branches, effectively evading detection by WASim, improving the hash rate to 102\% of its initial value while only increasing the file size by 19\%. 
This demonstrates that obfuscation can be viable in real-world scenarios, but it requires careful consideration of the specific use case.

Additionally, wasm-mutate presents potential as a tool for evading detection. 
Interestingly, in many cases, the performance overhead is negligible, and in some instances, wasm-mutate can even improve the hash rate. 
Although the file size overhead can be substantial, it can be mitigated by selectively applying transformations that do not significantly increase the file size.

While this paper focuses on crypto mining WebAssembly binaries, we have also included benign applications in our dataset. 
We find that benign applications can also be effectively obfuscated, and we have highlighted the differences between the different application categories. 
Although we have not extensively evaluated reverse engineering, the insights gained from this paper can likely be extended to benefit this domain as well.

\subsection{Limitations}

The dataset used in this paper covers a wide range of applications, but it is not exhaustive and only contains 22 applications.
This is primarily due to the constraints imposed by Tigress, which necessitates the use of open-source C projects compatible with the C99 standard.
Additionally, they must be compatible with \gls{CIL} so they can be parsed and merged into a single source file.
However, we were able to include all the prominent CryptoNight variants in the dataset.
This should provide extensive coverage of crypto mining malware, as several studies have found that in-the-wild cryptojacking implementations are all based on the CryptoNight algorithm~\cite{cabrera-arteaga-2022-webassemblydiversificationmalware, hilbig-2021-empiricalstudyreal}.
Moreover, with its diverse set of use cases, the dataset stands comparable to, and in some cases exceeds, the datasets of other obfuscation studies for WebAssembly~\cite{bhansali-2022-firstlookcode, cabrera-arteaga-2022-webassemblydiversificationmalware, Loose2023}.

The obfuscation methods have been tested on static detection methods and, despite their effectiveness in evading them, are unlikely to be equally effective for dynamic detection methods.
Although the transformations sometimes alter the observable behaviour of the program, as evidenced by the increase in native code, dynamic methods based on API calls or similar will observe the same behaviour with or without obfuscation.
However, as discussed in Section \ref{sec:background-detecting-drive-by-mining}, dynamic methods are complex to set up, can impose a significant performance overhead, and are not widely used in practice.
Therefore, we believe that the static detection methods that we evaluated are the most relevant for the purposes of this paper.

We could not extract the native code compiled by the browser, even after consulting the V8 developers.
This is unfortunate, as exploring the semantic differences in the native code after obfuscation would have provided valuable insights.
Still, we were able to extract the size of the native code, which indicates how obfuscation potentially affects the resulting native code and whether it is optimized away or not.
While we find this useful, we acknowledge that this is a limitation of our research.

\section{Future Research}

Although this paper has made significant contributions towards understanding WebAssembly obfuscation, there are still avenues for future research.
One crucial direction is the development of more robust detection methods that can effectively counter the obfuscation methods explored in this paper.
A promising solution is to preprocess the WebAssembly binaries by de-obfuscating them.

Moreover, the need for improved detection methods, irrespective of their obfuscation resistance, cannot be overstated.
In this paper, we discovered that more than half (four out of seven) of the detection methods implemented were unable to detect the crypto miners, even before obfuscation.

Another avenue for future research is investigating more advanced obfuscation techniques and possibly developing novel ones designed specifically for WebAssembly.
Obfuscation methods tailored to WebAssembly, leveraging its unique features such as the stack machine architecture, would likely be even more effective than the methods explored in this paper.
In practice, this could be implemented as optimization passes for Binaryen, similar to what was done for OLLVM.
Moreover, combining several obfuscation transformations, possibly at multiple abstraction levels, merits further investigation.

Finally, there is extensive research on cryptojacking but not on other malicious use cases for WebAssembly.
WebAssembly can also be used for other malicious purposes, like tech support scams, browser exploits, and script-based keyloggers~\cite{darkside-of-wasm}, and it has been used for hacking games~\cite{webassembly-2023-git-issue, wasm-2023-hacking}.
The full potential of WebAssembly for malicious purposes has not been extensively explored yet, and we believe that this is an exciting direction for future research.
Besides malicious use cases, obfuscating benign programs to prevent reverse engineering is another interesting direction for future research.

% ----------
% CONCLUSION 
% ----------

\section{Conclusion} \label{ch:conclusion}

In this paper, we have conducted a comprehensive evaluation of code obfuscation techniques for WebAssembly, covering multiple abstraction levels. 
This represents the most extensive assessment of WebAssembly obfuscation to date. We have demonstrated the effectiveness of obfuscation in producing dissimilar WebAssembly binaries and analyzed the impact on the resulting native code. 
The results show that obfuscation can successfully evade state-of-the-art cryptojacking detectors, although the effectiveness largely depends on the specific obfuscation transformation, detection method, and crypto mining algorithm employed.
Despite the observed performance overheads, we have demonstrated how obfuscation can be used to evade detection with justifiable overhead in real-world scenarios through careful selection of transformations.
These findings are significant for researchers and academics.
Our findings offer insights into which transformations are most effective in evading detection, which can inform the development of more robust detection methods. 
The novel obfuscation method introduced, along with the extensive dataset of obfuscated WebAssembly binaries, provides a solid foundation for future research in this domain as WebAssembly continues to gain widespread adoption across diverse platforms and use cases.

\bibliographystyle{plain}
\bibliography{references}

\end{document}